\documentclass[twocolumn]{aastex63}

\usepackage{graphicx} 
\usepackage{multirow}
\usepackage{booktabs}
\usepackage{colortbl} 
\usepackage{xcolor} 
\usepackage{amsmath, amsthm, amssymb, amsfonts}
\usepackage{array}
\usepackage{mathptmx}
\usepackage{hyperref}

\begin{document}

\title{Hadronic Processes in Advection-Dominated Accretion Flow as the Origin of TeV Excesses in BL Lac Objects}

\correspondingauthor{Jin Zhang}
\email{j.zhang@bit.edu.cn}

\author[0000-0003-2547-1469]{Ji-Shun Lian}
\affiliation{School of Physics, Beijing Institute of Technology, Beijing 100081, People's Republic of China; j.zhang@bit.edu.cn}

\author[0000-0002-3883-6669]{Ze-Rui Wang}
\affiliation{College of Physics and Electronic Engineering, Qilu Normal University, Jinan 250200, People’s Republic of China}

\author[0000-0003-3554-2996]{Jin Zhang\dag}

\affiliation{School of Physics, Beijing Institute of Technology, Beijing 100081, People's Republic of China; j.zhang@bit.edu.cn}

\begin{abstract}

The spectral energy distributions (SEDs) of certain BL Lac objects (BL Lacs) exhibit an additional hard $\gamma$-ray component in the TeV energy range that surpasses the predictions of the one-zone leptonic jet model. The origin of this excess emission remains unclear. In this study, we selected five BL Lacs whose SEDs display a very hard intrinsic spectrum in the TeV band and successfully reproduced their broadband SEDs using a two-zone lepto-hadronic model. Within this framework, the emission observed in the optical, X-ray, GeV $\gamma$-ray, and sub-TeV $\gamma$-ray bands is modeled using the synchrotron and synchrotron self-Compton radiation processes of the relativistic electrons in the jets. Meanwhile, the TeV excess is attributed to $\gamma$-ray emission resulting from the photomeson ($p\gamma$) process via $\pi^0$ decay occurring within advection-dominated accretion flows (ADAFs). This scenario requires a hard proton spectrum with a spectral index of $p \sim 1.6-1.7$ and a cutoff energy ranging from 30 to 90 TeV, as well as a relatively large ADAF radius. Such hard proton spectra suggest that the dominant acceleration mechanisms are likely magnetic reconnection and/or stochastic acceleration processes within ADAFs. Additionally, the emission from the cascaded electrons results in a bump in the keV--MeV band; however, it is overwhelmed by the jet emission. Although the hadronuclear ($pp$) process cannot be entirely ruled out, it would necessitate an even harder proton spectrum and a higher cutoff energy compared to the $p\gamma$ process, making it a less favorable explanation for the observed TeV excess. 

\end{abstract}


\section{Introduction} 

Blazars are a subclass of active galactic nuclei (AGNs) characterized by relativistic jets that are aligned closely with the observer's line of sight, typically within a few degrees ($<5\degr$, \citealt{1993ApJ...407...65G, 1995PASP..107..803U, 2019ARA&A..57..467B}). Blazars are further classified into flat spectrum radio quasars (FSRQs) and BL Lac objects (BL Lacs) on the basis of their optical spectral features; FSRQs exhibit prominent emission lines, while BL Lacs have weak or no emission lines \citep{1991ApJ...374..431S, 2003ApJ...585L..23F, 2012MNRAS.420.2899G}. According to the Fermi-LAT catalog data, BL Lacs generally demonstrate lower luminosity and harder spectra compared to FSRQs \citep{2017ApJS..232...18A}. The jets of these two types of blazars are believed to differ in both composition and radiation efficiency \citep{2014Natur.515..376G, 2014ApJ...788..104Z}, a distinction that is intrinsically linked to the varying accretion rates of their central supermassive black holes (SMBHs, \citealt{2009MNRAS.396L.105G, 2013MNRAS.431.1914G, 2014Natur.515..376G, 2022ApJ...925...97P}). Within the framework of unified models for radio-loud AGNs, BL Lacs and FSRQs are associated with Fanaroff-Riley type I (FR I) and type II (FR II) radio galaxies \citep{1995PASP..107..803U}, respectively. These two types of radio galaxies are characterized by distinct accretion modes due to the different accretion rates \citep{2014MNRAS.440..269M, 2015MNRAS.447.1184F}; the hot mode of a radiatively inefficient accretion flow (RIAF) for FR I radio galaxies and the cold mode of a standard thin accretion disk for FR II radio galaxies \citep{2016CRPhy..17..594D, 2018Galax...6..116R}. Therefore, BL Lacs are likely to host a RIAF. 

Blazars represent one of the primary extragalactic $\gamma$-ray emitting sources, especially in the TeV band, with the majority of detected TeV sources being BL Lacs\footnote{\url{http://tevcat2.uchicago.edu/} \citep{2008ICRC....3.1341W}}. The broadband spectral energy distributions (SEDs) of BL Lacs typically exhibit a double-peaked structure, which is commonly attributed to synchrotron radiation and synchrotron self-Compton (SSC) processes of relativistic electrons (e.g., \citealt{1992ApJ...397L...5M, 2010MNRAS.402..497G, 2010MNRAS.401.1570T, 2012ApJ...752..157Z, 2014MNRAS.439.2933Y}). In general, this one-zone SSC model provides an adequate description of the BL Lac SEDs during quiescent states, such as those of BL Lacs Mrk 421 \citep{2011ApJ...736..131A} and H2356--309 \citep{2010A&A...516A..56H}. However, when sources exhibit high-flux activity in the X-ray and/or $\gamma$-ray bands or display a hard $\gamma$-ray spectrum in the TeV regime, more complex multi-zone models are often required to accurately reproduce the observed broadband emission (e.g., \citealt{2005ApJ...630..130B, 2007A&A...470..475A}). With a multiyear observational campaign conducted by the MAGIC Collaboration on 10 extreme high-frequency-peaked BL Lacs, \citet{2020ApJS..247...16A} found that while the one-zone SSC model can still fit the data, it requires a critically low magnetization. In contrast, a spine-layer emission model offers a satisfactory solution to this magnetization problem. Furthermore, recent multi-wavelength photometric and polarization studies provide growing evidence in favor of multiple emission zones (e.g., \citealt{2022Natur.611..677L, 2025A&A...695A.217M}). Additionally, the one-zone leptonic model encounters difficulties in accounting for the detection of a neutrino event associated with the BL Lac TXS 0506+056 \citep{2018Sci...361.1378I}. Therefore, hybrid leptonic--hadronic jet models have been proposed to explain such  multi-messenger observations \citep{2018ApJ...863L..10A, 2018ApJ...864...84K, 2019ApJ...886...23X}.

The radio-quiet Seyfert galaxy NGC 1068 has been confirmed as an astrophysical neutrino emitter \citep{2022Sci...378..538I}, with the observed neutrinos likely originating from high-energy protons accelerated in the corona (e.g., \citealt{2022ApJ...941L..17M, 2024ApJ...972...44D, 2024ApJ...974...75F, 2024PhRvD.109j1306M}). Similar mechanisms have also been proposed to explain the origin of $\gamma$-ray emission and to predict the potential neutrino flux from other Seyfert galaxies, including NGC 4151, NGC 4945, and Circinus galaxy \citep{2024ApJ...961L..34M}, as well as the detected neutrino emission from the blazar TXS 0506+056 \citep{2025ApJ...984...54K, 2025ApJ...980..255Y}. Moreover, \citet{2015ApJ...806..159K} suggested that low-luminosity AGNs (LLAGNs) could generate high-energy neutrino emission via $pp$ and/or $p\gamma$ interactions, where protons are accelerated stochastically by turbulent processes within RIAFs. These findings imply that disk-corona systems are likely also sources of $\gamma$-ray emission due to the presence of accelerated high-energy particles, where the disk-corona system includes RIAFs.

Some studies have suggested that the accretion mode of BL Lacs is the optically thin advection-dominated accretion flows (ADAFs, \citealt{2002ApJ...579..554W,2023MNRAS.526.4079C, 2024ApJ...967..104Z}), which represents one of the most extensively studied RIAF models \citep{1994ApJ...428L..13N,2007ASPC..373...95Y,2014ARA&A..52..529Y}. Within ADAFs, nonthermal protons can be accelerated by magnetic reconnections \citep{2013ApJ...773..118H, 2019ApJ...886..122C, 2021MNRAS.506.1128S},  magnetohydrodynamic (MHD) turbulence \citep{2006ApJ...647..539B, 2008ApJ...681.1725S, 2015ApJ...806..159K, 2021NatCo..12.5615K}, or shocks \citep{1991PhRvL..66.2697S, 2012ApJ...754..148T}. In this paper, we investigate the radiation of high-energy protons within ADAFs to provide an explanation for the TeV $\gamma$-ray excess observed in five BL Lacs, a phenomenon defined in reference to the one-zone SSC model. The broadband SED data for the selected sources are presented in Section~\ref{sec_sample}, while the details of the proposed model are outlined in Section~\ref{sec_Model}. The model fitting strategies and corresponding results are given in Section~\ref{sec_result}. Discussion of fitting results and a summary are provided in Sections~\ref{sec_dis} and \ref{sec_Summary}, respectively.

\section{TeV-Excess Sources} \label{sec_sample}

\begin{figure*}
    \centering
    \includegraphics[angle=0,width=0.49\textwidth]{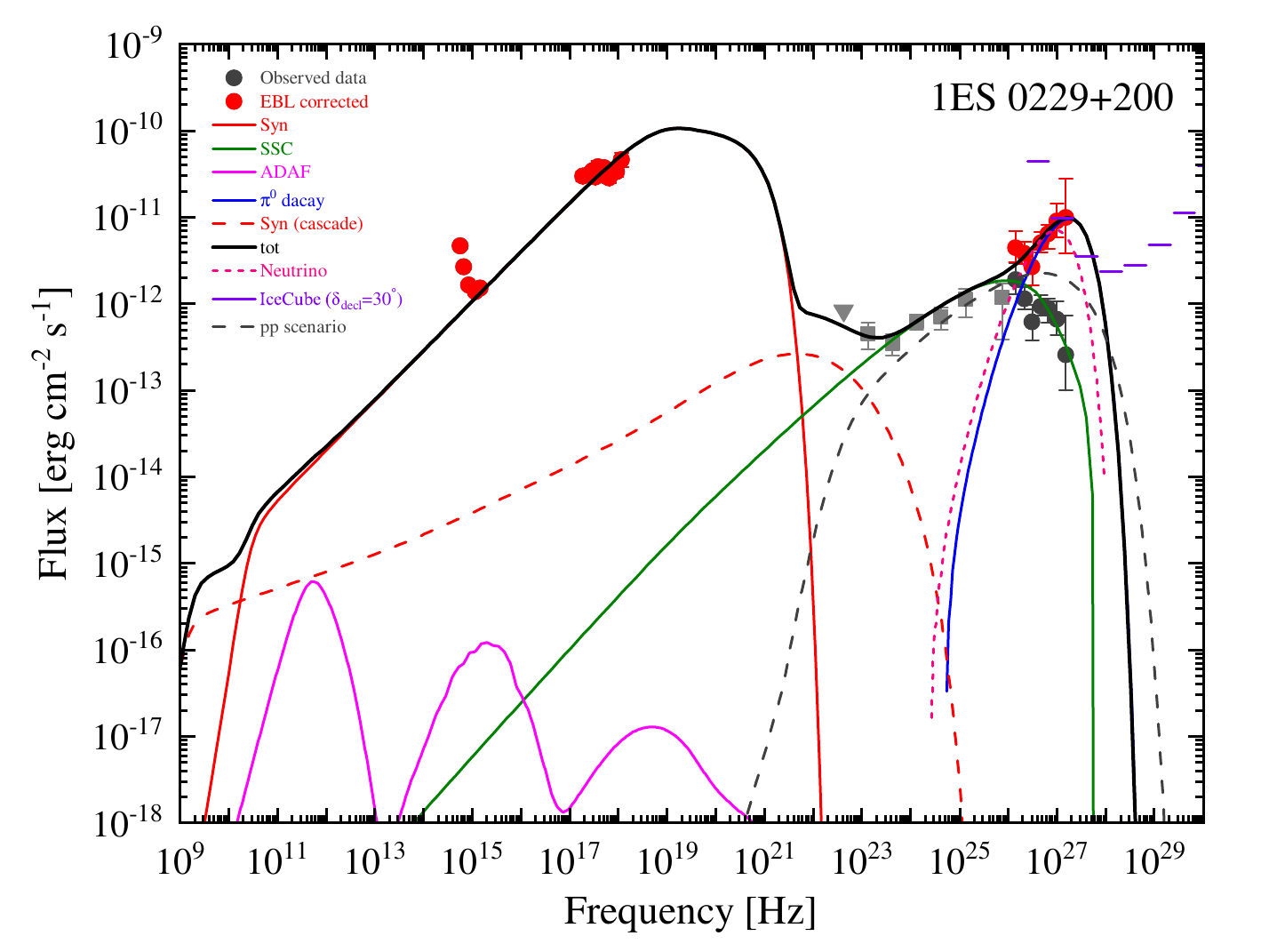}
    \includegraphics[angle=0,width=0.49\textwidth]{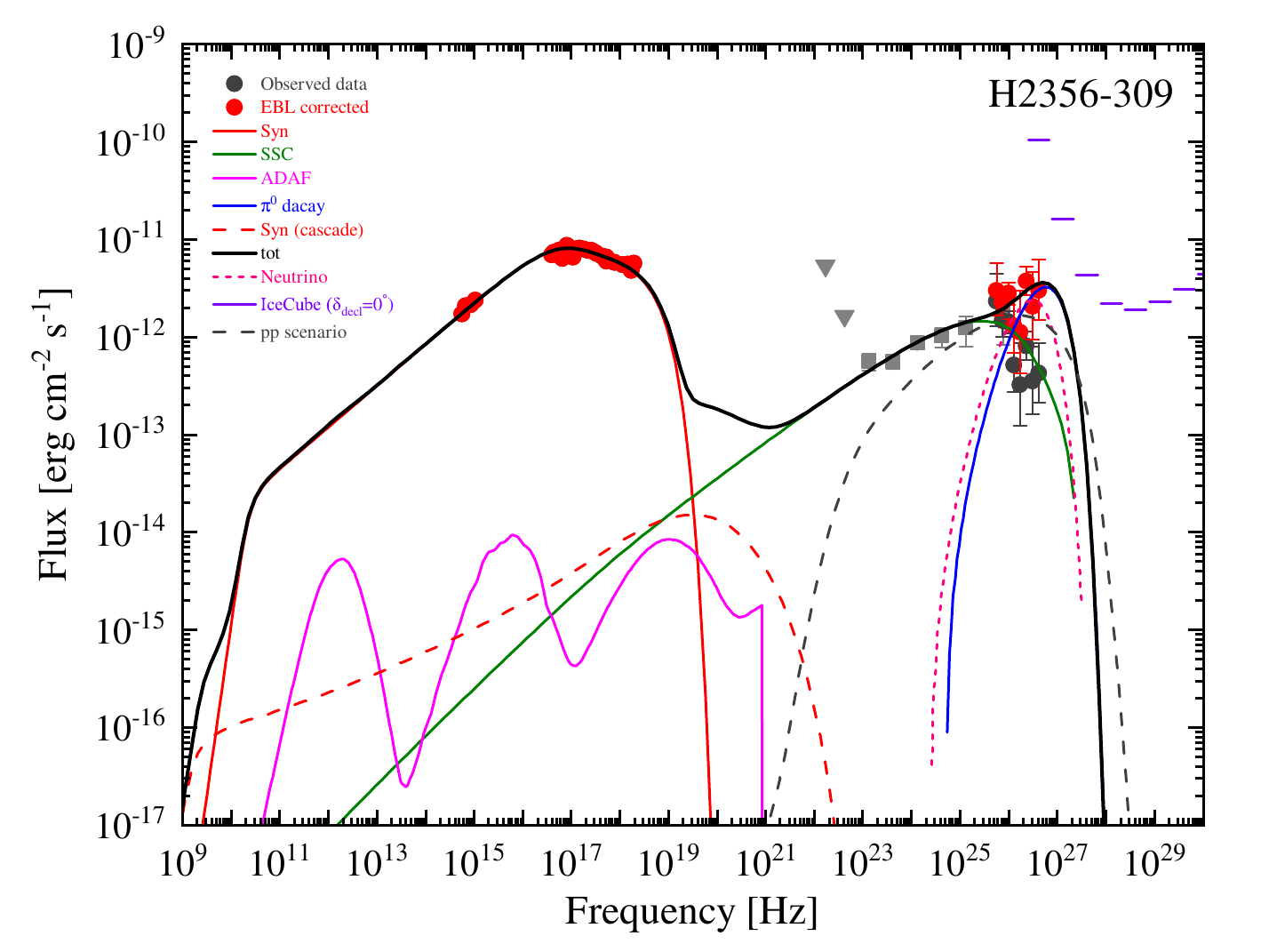}     
    \includegraphics[angle=0,width=0.49\textwidth]{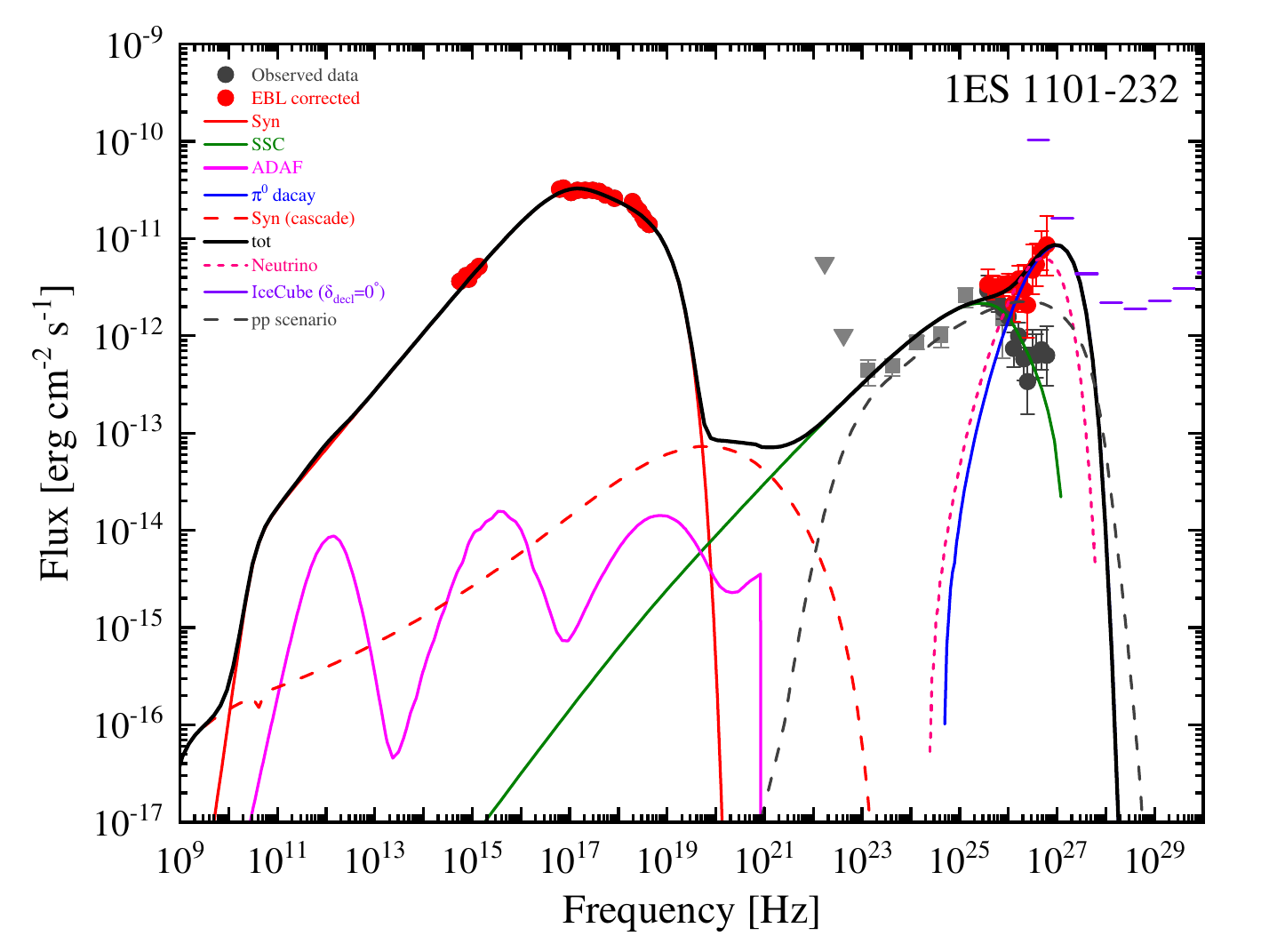}
    \includegraphics[angle=0,width=0.49\textwidth]{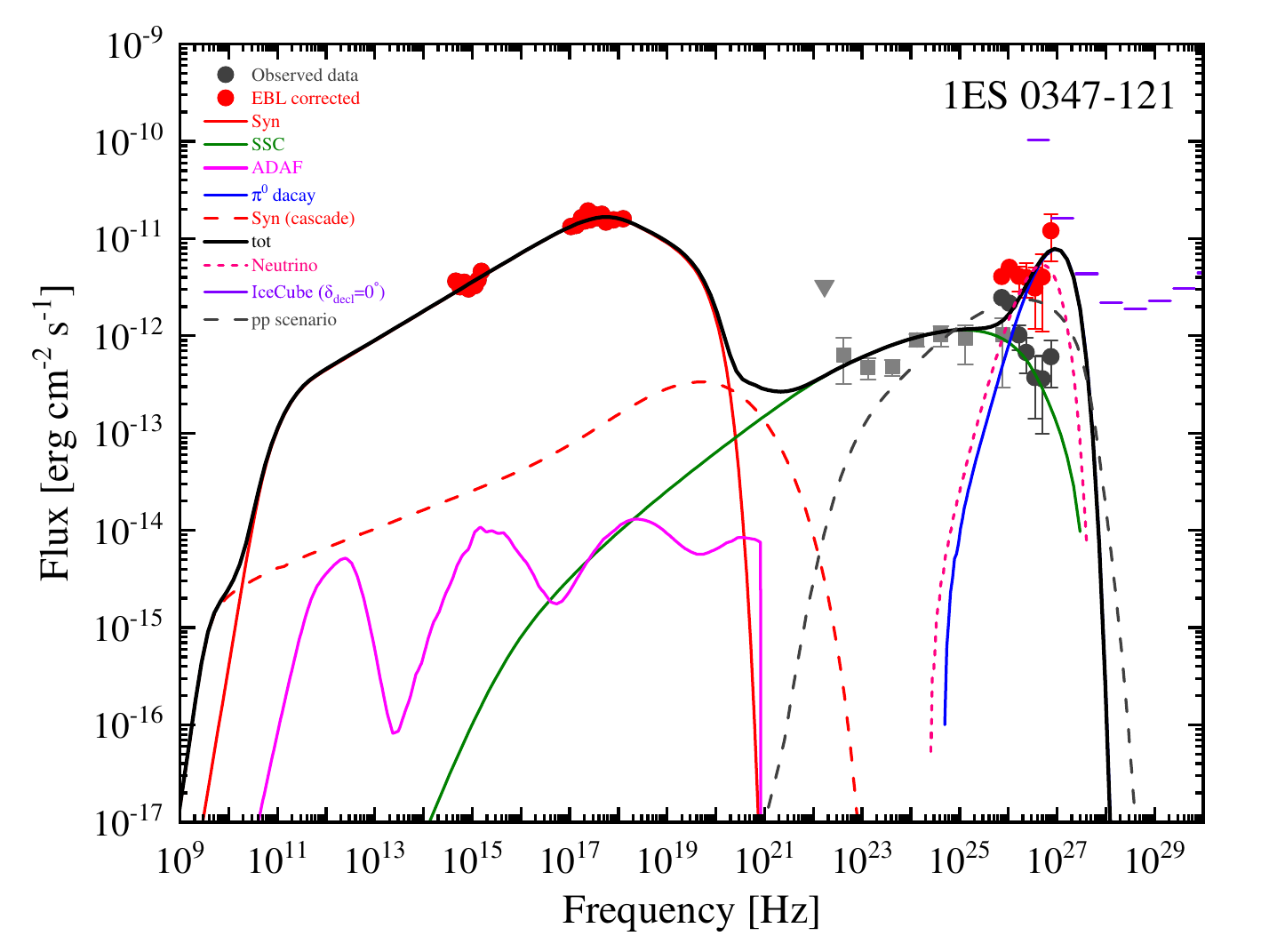}    
    \includegraphics[angle=0,width=0.49\textwidth]{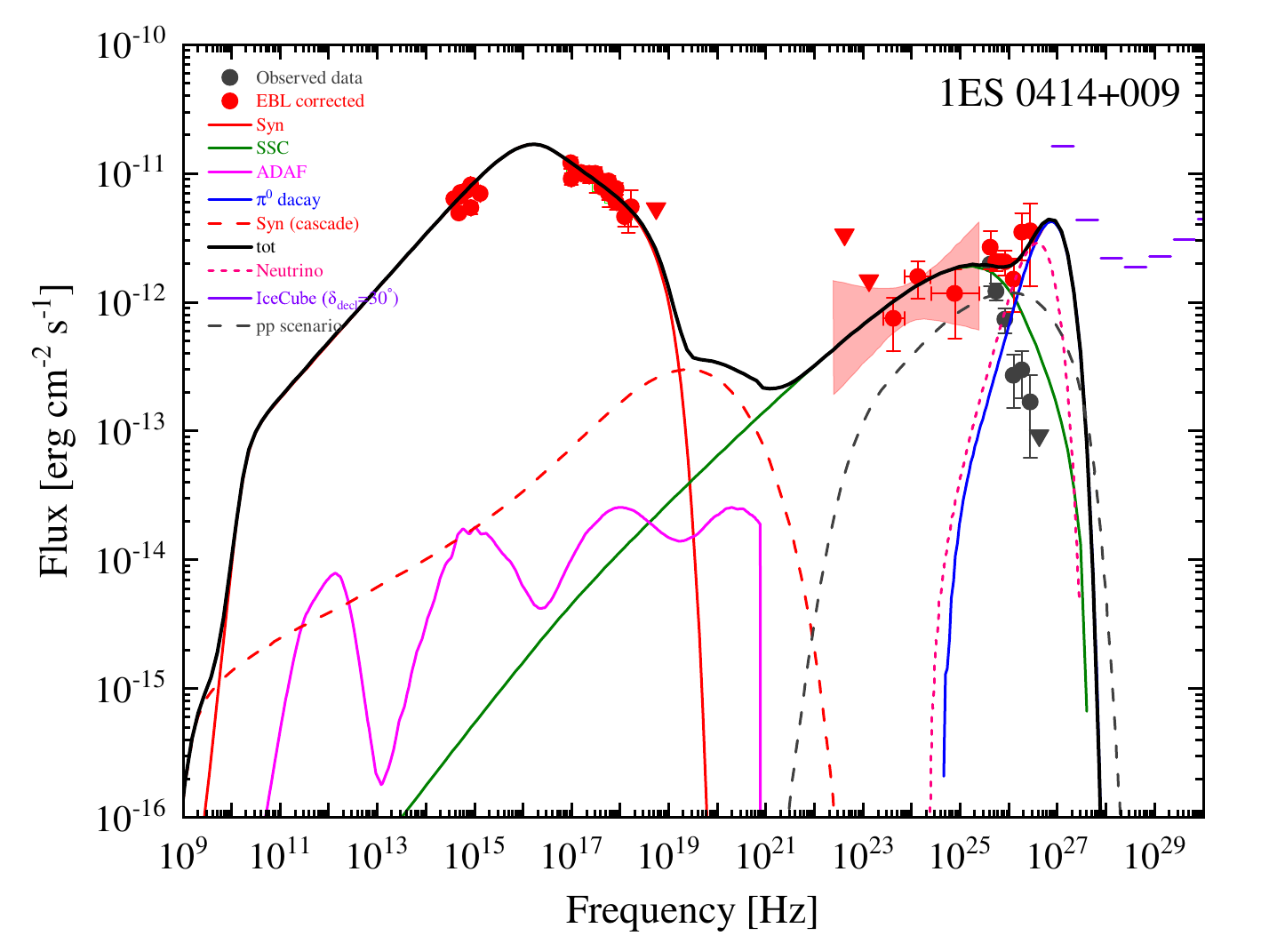}
    \caption{Broadband SEDs of the five BL Lacs with fitting results. The observational data, represented by scattered points, are referenced to Section \ref{sec_sample}. Gray circles and triangles denote data in the observer frame, whereas the corresponding red symbols indicate data corrected using the EBL model of \citet{2022ApJ...941...33F}. The black solid lines represent the sum of each component emission for the sources, including the jet synchrotron radiation (red solid lines), the jet SSC radiation without EBL absorption (green solid lines), the ADAF spectrum (magenta solid lines), the $\gamma$-ray emission from the $\pi^0$ decay in the $p\gamma$ process (blue solid lines) and in the $pp$ process (gray dashed lines), and the radiation from cascaded electrons (red and green dashed lines) produced by photomeson and Bethe--Heitler processes in the ADAF.
    The differential sensitivity curves (violet short horizontal lines) of IceCube are taken from \citet{2019EPJC...79..234A}.}    
    \label{Fig: SED}
\end{figure*}

This study includes five BL Lac SEDs that exhibit a significant TeV excess component, where the TeV excess is revealed based on the one-zone SSC model fitting results. Among these, four sources\footnote{Mrk 421 is excluded due to its significant variability across multiple wavelengths, particularly in the TeV band.} are drawn from \citet{2022ApJ...925L..19C}, with the TeV excess attributed to the \textit{p$\gamma$} process occurring within the jets. The selected SEDs of these five sources are illustrated in Figure \ref{Fig: SED}, with the intrinsic TeV emission corrected based on the extragalactic background light (EBL) model provided by \citet{2022ApJ...941...33F}. Additionally, based on the long-term light curves presented in the Fermi-LAT 14-year Source Catalog (4FGL-DR4)\footnote{\url{https://fermi.gsfc.nasa.gov/ssc/data/access/lat/14yr_catalog/}}, no significant variability is detected in the GeV band for these sources. Therefore, we also include the 14-year average spectra from the 4FGL-DR4 for these sources (with the exception of 1ES 0414+009) in Figure \ref{Fig: SED} to constrain the model parameters. Further details regarding the selected sources and their SEDs are outlined below.

\begin{itemize}

\item \textit{1ES 0229+200.} A typical extreme high-synchrotron-peaked BL Lac at a redshift of $z = 0.1396$ \citep{2005ApJ...631..762W}, characterized by an extremely hard spectrum in both the X-ray and TeV $\gamma$-ray bands. Its TeV $\gamma$-ray emission was first detected by the High Energy Stereoscopic System (H.E.S.S.) telescopes in 2005 and 2006 \citep{2007A&A...475L...9A}, with the spectrum extending up to 15 TeV and exhibiting a hard photon spectral index of $\Gamma_{\gamma}\sim2.5$ \citep{2007A&A...475L...9A}. As displayed in Figure \ref{Fig: SED}, the TeV $\gamma$-ray data were obtained from observations with the H.E.S.S. telescopes in 2006 \citep{2007A&A...475L...9A}, while the optical and X-ray data, collected in 2008, were provided by Swift-UVOT and Swift-XRT, respectively \citep{2010MNRAS.401.1570T}.

\item \textit{H2356--309.} A high-synchrotron-peaked BL Lac located at a redshift of $z = 0.165$ \citep{1991AJ....101..821F}. H2356--309 is a candidate neutrino emitter situated within the error circles of three IceCube events, suggesting that the TeV $\gamma$-ray emission could originate from hadronic processes \citep{2015EPJC...75..273S}. The quasi-simultaneous broadband SED, as illustrated in Figure \ref{Fig: SED}, is taken from \citet{2010A&A...516A..56H}. It includes optical/UV and X-ray observations conducted by XMM-Newton on 15 June 2005, as well as the time-averaged very high energy (VHE) $\gamma$-ray spectra obtained from observations with the H.E.S.S. telescopes in 2005.

\item \textit{1ES 1101--232.} A high-synchrotron-peaked BL Lac located at a redshift of $z = 0.186$ \citep{1994ApJS...93..125F}. Its VHE $\gamma$-ray emission was first detected by the H.E.S.S. telescopes during observations conducted in 2004 and 2005 \citep{2007A&A...470..475A}. The source displayed an exceptionally hard spectrum, with a photon spectral index of $\Gamma_{\gamma} \sim 1.5$ in the energy range of 0.23--4 TeV, and no significant variability in the VHE $\gamma$-ray flux was observed across any time scale. The quasi-simultaneous SED, as illustrated in Figure \ref{Fig: SED}, incorporates X-ray and optical data obtained from XMM-Newton observations on 8 June 2004, as well as H.E.S.S. data collected between 5 and 16 March 2005.

\item \textit{1ES 0347--121.} An extremely high-synchrotron-peaked BL Lac located at a redshift of $z = 0.188$ \citep{2005ApJ...631..762W}. The VHE $\gamma$-ray emission from this source was first detected by the H.E.S.S. telescopes in 2006 \citep{2007A&A...473L..25A}. The possibility of a hadronic jet origin for its VHE emission has been investigated in \citet{2015MNRAS.448..910C}. The H.E.S.S. observational data, together with the simultaneous X-ray and UV/optical data obtained from Swift and the Automatic Telescope for Optical Monitoring, are retrieved from \citet{2007A&A...473L..25A} and used to construct the broadband SED, as shown in Figure~\ref{Fig: SED}. 

\item \textit{1ES 0414+009.} One of the most distant TeV BL Lacs detected to date, located at a redshift of $z = 0.287$ \citep{1991AJ....101..818H}. Its VHE $\gamma$-ray emission was first detected by the H.E.S.S. telescopes during observations conducted between October 2005 and December 2009 \citep{2012A&A...538A.103H}. The source was also detected by VERITAS between January 2008 and February 2011 \citep{2012ApJ...755..118A}. Broadband data, including optical observations from the Michigan-Dartmouth-MIT Observatory, X-ray data from Swift, GeV observations from Fermi-LAT, and the VERITAS spectrum, indicate that a homogeneous one-zone leptonic model is insufficient to accurately describe the broadband SED of 1ES 0414+009. In contrast, a lepto-hadronic model provides a better fit to the data \citep{2012ApJ...755..118A}. The selected broadband SED of 1ES 0414+009, as shown in Figure~\ref{Fig: SED}, is obtained from \citet{2012A&A...538A.103H}.

\end{itemize}

\section{model} \label{sec_Model}

The one-zone leptonic jet model is commonly used to fit the observed broadband SEDs of TeV-emitting BL Lacs (e.g., \citealt{2010MNRAS.401.1570T, 2012ApJ...752..157Z}). In this work, a two-zone lepto-hadronic model is employed to account for the broadband SEDs of the five BL Lacs. The observed optical, X-ray, and GeV $\gamma$-ray emission is attributed to the synchrotron and SSC processes of the relativistic electrons within the jets, following a similar approach to that in \citet{2010MNRAS.401.1570T} and \citet{2012ApJ...752..157Z}. Meanwhile, the TeV excess emission, which exceeds the predictions of the one-zone SSC model, is interpreted as the decay of neutral pions produced via $p\gamma$ interactions within the ADAFs.

\subsection{Leptonic Radiation within the Jet}

The emission region is modeled as a sphere with a radius of $R_{\rm b}$ and magnetic field strength of $B_{\rm jet}$, as adopted in previous studies (e.g., \citealt{2012ApJ...752..157Z, 2017ApJ...842..129C, 2021RAA....21..103X, 2022PhRvD.105b3005W}). The size of the emission region is derived with the variability timescale ($\Delta t$) using the relation $R_{\rm b} = c \Delta t\delta_{\rm D}/(1 + z)$, where $\delta_{\rm D}$ denotes the Doppler factor, and $c$ is the speed of light. We assume $\delta_{\rm D}=\Gamma$ for these BL Lacs, where $\Gamma$ is the bulk Lorenz factor of the emission region. The energy distribution of relativistic electrons is described by a broken power-law function, defined by a normalization constant $N_{\rm e,0}$, a break energy $\gamma_{\rm e,b}$, and power-law indices $p_1$ and $p_2$ below and above the break energy, respectively, within the energy range [$\gamma_{\rm e,min}$, $\gamma_{\rm e,max}$]. The synchrotron and SSC processes of relativistic electrons, along with the effects of synchrotron self-absorption and Klein--Nishina, are incorporated into the leptonic model calculations. 

\subsection{Lepto-hadronic Radiation within the ADAF}

BL Lacs are believed to harbor an ADAF in the central region near the SMBHs. ADAF is characterized by being geometrically thick, optically thin, and having low radiative efficiency (e.g., \citealt{1982Natur.295...17R, 1977ApJ...214..840I, 1994ApJ...428L..13N, 2007ASPC..373...95Y, 2014ARA&A..52..529Y}). Within ADAFs, thermal electrons and ions lose energy through synchrotron emission, bremsstrahlung, and inverse Comptonization of seed photons generated by these two radiative processes, thereby producing the ADAF emission spectrum across the electromagnetic spectrum from radio to hard X-ray bands. Using the parameter values listed in Table \ref{table_SED}, we numerically compute the global structure and dynamics of the flows following the methods outlined in \citet{2000ApJ...537..236Y, 2003ApJ...598..301Y} to calculate the ADAF spectrum. For further details, please refer to \citet{2005ApJ...620..905Y, 2007ApJ...659..541Y} and \citet{2014MNRAS.438.2804N}. The parameters include the SMBH mass ($M_{\rm BH}$), the power-law index of radial variation ($s$), the viscosity parameter ($\alpha$), the ratio of gas pressure to total pressure ($\beta$), the adiabatic index ($\gamma_{\rm adi}$), the outer radius of the ADAF ($R_{\rm o}$), the fraction of the turbulent energy that heats the electrons ($\delta_{\rm ADAF}$), and the accretion rate ($\dot{m}$).

We hypothesize the existence of relativistic protons accelerated within the ADAFs of these BL Lacs. The energy distribution of these protons is assumed to follow an exponential cutoff power-law function, characterized by a spectral index $p$ and a cutoff energy $\varepsilon_{\rm p,cut}$, within the energy range [$\varepsilon_{\rm p,min}$, $\varepsilon_{\rm p,max}$], i.e.,  
\begin{equation}
N_{\mathrm{p}}\left(\varepsilon_{\mathrm{p}}\right)=N_{\rm p,0}\left(\frac{\varepsilon_{\mathrm{p}}}{1 \mathrm{eV}}\right)^{-p} \exp \left(-\frac{\varepsilon_{\mathrm{p}}}{\varepsilon_{\mathrm{p}, \text {cut}}}\right).
\end{equation}
These protons interact with photons emitted by the ADAF through the $p\gamma$ process. Both photomeson production and the Bethe--Heitler process ($p \gamma \rightarrow p + e^{+} + e^{-}$) are taken into account. It is assumed that the relevant hadronic processes occur within a spherical ADAF of radius $R_{\rm o}$. The viscosity parameter of the ADAFs is set to $\alpha \sim 0.3$; therefore, the accretion flow is quasi-spherical rather than toroidal \citep{1998tbha.conf..148N}.

The photomeson process is calculated using a semi-analytical method. The proton energy loss rate is given by 
\begin{equation}
t_{p \gamma}^{-1}\left(\varepsilon_p\right)=\frac{c}{2 \gamma_p^2} \int_{\bar{\varepsilon}_{\text {th }}}^{\infty} d \bar{\varepsilon} \sigma_{p \gamma}(\bar{\varepsilon}) \kappa_{p \gamma}(\bar{\varepsilon}) \bar{\varepsilon} \int_{\bar{\varepsilon} / 2 \gamma_p}^{\infty} d \varepsilon \varepsilon^{-2} n_{\varepsilon},
\end{equation}
where $\varepsilon_{\rm p}$ and $\varepsilon$ are the proton and photon energies in the ADAF frame, respectively; $\bar{\varepsilon}$ is the photon energy in the proton frame; $\sigma_{p\gamma}$ and $\kappa_{p\gamma}$ are the photomeson cross section and the proton inelasticity, respectively (for details, see \citealt{1968PhRvL..21.1016S, 1992ApJ...400..181C, 2003ApJ...586...79A, 2008A&A...485..623R, 2022ApJ...925L..19C}). Relativistic protons lose energy through pion production, which includes single-pion and multi-pion branching processes as described by   
\begin{equation}
    p+\gamma \rightarrow \begin{cases}\mathrm{p}+\mathrm{a} \pi^0+\mathrm{b}\left(\pi^{+}+\pi^{-}\right), \\
     \mathrm{n}+\pi^{+}+a \pi^0+b\left(\pi^{+}+\pi^{-}\right), \end{cases}
\end{equation}
where $a$ and $b$ are the pion multiplicities \citep{2008A&A...485..623R}. Subsequently, the produced pions decay further as follows:
\begin{equation}\label{eq. neutino}
\begin{array}{ll}
\pi^{+} \rightarrow \mu^{+}+v_\mu, & \mu^{+} \rightarrow \mathrm{e}^{+}+v_{\mathrm{e}}+\bar{v}_\mu, \\
\pi^{-} \rightarrow \mu^{-}+\bar{v}_\mu, & \mu^{-} \rightarrow \mathrm{e}^{-}+\bar{v}_{\mathrm{e}}+v_\mu, \\
\pi^0 \rightarrow \gamma \gamma.
\end{array}
\end{equation}
The observed TeV excess in the broadband SEDs of the five BL Lacs is attributed to the decay of $\pi^0$ mesons.

The potential influence of hadron-triggered electromagnetic cascade processes within the ADAF is also investigated. The cascade processes involve electrons that are injected through $\gamma \gamma$ annihilation, the photopion process, and the Bethe--Heitler process. The Bethe--Heitler process is computed using a semi-analytical method in the framework established by \citet{2008PhRvD..78c4013K}. The distribution of cascaded electrons is determined through a time-independent approach by iteratively solving the isotropic Fokker-Planck equation under equilibrium conditions (\citealt{2013ApJ...768...54B}):
\begin{equation}
\frac{\partial}{\partial \gamma_{\mathrm{e}}}\left(\dot{\gamma}_{\mathrm{e}} N_{\mathrm{e}}^{\mathrm{cas}}\right)=Q_{\mathrm{e}}\left(\gamma_{\mathrm{e}}\right)+\dot{N}_{\mathrm{e}}^{\gamma \gamma}\left(\gamma_{\mathrm{e}}\right)+\dot{N}_{\mathrm{e}}^{\mathrm{esc}}\left(\gamma_{\mathrm{e}}\right),
\end{equation}
where $Q_{\rm e}$ represents the total electron injection resulting from the Bethe--Heitler and photomeson processes, $\dot{N}^{\gamma \gamma}_{\rm e}$ denotes the electron injection from internal $\gamma \gamma$ interactions, $\dot{N}^{\rm esc}_{\rm e}$ accounts for the escape term, and $N^{\rm cas}_{\rm e}$ indicates the cascade electron distribution. The cascaded electrons lose energy primarily through synchrotron and SSC processes.

\section{Fitting Strategies and Results}\label{sec_result}

The broadband SEDs of the five BL Lacs are modeled using the synchrotron and SSC emission processes from a single population of relativistic electrons within the jets, as well as the radiation triggered by a single population of protons accelerated in ADAFs. 

Firstly, we use the leptonic jet radiation to account for the observed SEDs, while ignoring the TeV flux fitting. The leptonic jet model incorporates the following parameters: $B_{\rm jet}$, $\Delta t$, $\delta_{\rm D}$, $p_1$, $p_2$, $\gamma_{\rm e,min}$, $\gamma_{\rm e,b}$, $\gamma_{\rm e,max}$, and $N_{\rm e,0}$. Except for 1ES 0414+009, the same set of observational data was adopted for the remaining sources studied in \citet{2022ApJ...925L..19C}, where the same leptonic model was applied to reproduce the broadband SEDs of the five BL Lacs. Consequently, comparable values for the fitting parameters are obtained in this study. For 1ES 0414+009, the variability timescale is from \citet{2012ApJ...755..118A}. The derived parameter values are expected to be broadly consistent with those reported in previous studies of these five BL Lacs, where the one-zone synchrotron+SSC model was similarly applied to interpret the broadband SEDs (e.g., \citealt{2007A&A...473L..25A, 2007Ap&SS.309..487C, 2010A&A...516A..56H, 2010MNRAS.401.1570T, 2012ApJ...752..157Z}). A summary of all fitted parameters is provided in Table \ref{table_SED}, and the corresponding fitting results are illustrated in Figure \ref{Fig: SED}.

Secondly, the ADAF emission spectrum resulting from thermal electrons and ions is calculated based on the following parameters: $M_{\rm BH}$, $s$, $\alpha$, $\beta$, $\gamma_{\rm adi}$, $R_{\rm o}$, $\delta_{\rm ADAF}$, and $\dot{m}$. During the calculations, we fix the values of $s=0.3$, $\alpha=0.3$, $\beta=0.9$, $\gamma_{\rm adi}=1.5$, and $R_{\rm o} = 10^4R_{\rm S}$ ($R_{\rm S}$ is the Schwarzschild radius) for all sources. The values of $M_{\rm BH}$ for each source are listed in Table \ref{table_SED}, along with the corresponding references. The accretion rate is defined as $\dot{m}=\dot{M} / \dot{M}_{\mathrm{Edd}}$, where $\dot{M}$ represents the SMBH accretion rate, and $\dot{M}_{\mathrm{Edd}}=L_{\mathrm{Edd}} / \eta c^2$ denotes the Eddington accretion rate. Here, $L_{\rm Edd}$ refers to the Eddington luminosity, $\eta = \frac{L_{\mathrm{disk}}}{\dot{M} c^2}$ \citep{1992apa..book.....F, 1998tbha.conf..148N} is the radiative efficiency, and $L_{\rm disk}$ corresponds to the disk luminosity. Therefore, the accretion rate can also be expressed as $\dot{m} = \frac{L_{\mathrm{disk}}}{L_{\mathrm{Edd}}}$. Since the directly measured $L_{\rm disk}$ values for these five BL Lacs are unavailable, the luminosity at $10^{15}$ Hz from the leptonic jet model fitting curve is used as an approximation for $L_{\rm disk}$ to compute the corresponding accretion rates\footnote{In this case, the accretion rates of the sources may be overestimated, leading to an overestimation of the luminosity of ADAFs. This, in turn, results in a lower proton number density or a softer proton spectral index when modeling the observational data. However, this effect is negligible. A decrease in ADAF luminosity by one order of magnitude would require only a minor adjustment of approximately 0.1 in the proton spectral index.}, as summarized in Table \ref{table_SED}.
At this stage, only $ \delta_{\rm ADAF} $ remains as a free parameter. Once $ \delta_{\rm ADAF} $ is fixed, the ADAF spectrum of thermal electron and ion emission can be derived. The parameter $ \delta_{\rm ADAF} $ is generally assumed to lie within the range of 0.1 to 0.5 \citep{2014ARA&A..52..529Y}. We primarily set it to 0.5, although minor adjustments may be made to optimize the ADAF photon fields for the p$\gamma$ process.

Subsequently, the spectrum of the relativistic proton radiations through the photomeson and cascade processes is modeled using the parameters: $R_{\rm o}$, the magnetic field strength ($B_{\rm ADAF}$) within the ADAF, and the proton distribution parameters ($p$, $\varepsilon_{\rm p,min}$, $\varepsilon_{\rm p,cut}$, $\varepsilon_{\rm p,max}$, $N_{\rm p,0}$). It should be noted that the target photons involved in the $p\gamma$ process are considered exclusively from the ADAF spectrum. The values of $R_{\rm o}$ have been fixed as previously described. The magnetic field strength within ADAFs, from the inner to the outer regions, is determined by the following relation \citep{2014ARA&A..52..529Y}
\begin{equation}
\begin{gathered}
    B_{\rm ADAF} \approx 6.5 \times 10^8(1+\beta)^{-1 / 2} \alpha^{-1 / 2} \\
    M^{-1 / 2}_{\mathrm{BH}} \dot{m}^{1 / 2} r^{-5 / 4+s / 2}\ \mathrm{G},
\end{gathered}
\end{equation}
where $r \equiv R / R_{\rm S} $, and $R$ is the radius within $R_{\rm S} \leq R \leq R_{\rm o}$.
Based on this equation, we assume a uniform magnetic field with an average value of $ B_{\rm ADAF}=$1 G, which is also consistent with results given in other studies (e.g., \citealt{2003ApJ...598..301Y, 2015ApJ...806..159K, 2019PhRvD.100h3014K, 2021NatCo..12.5615K}). Accordingly, $ B_{\rm ADAF} = 1 $ G is applied to the five BL Lacs.

Finally, based on the results of the leptonic jet fitting, the proton distribution parameters and the value of $\delta_{\rm ADAF}$ are adjusted to model the TeV excess component in the SEDs of five BL Lacs. Our numerical results are presented in Figure \ref{Fig: SED}, and the corresponding model parameters are summarized in Table \ref{table_SED}. It is evident that the model well reproduces the observational data. However, it should be noted that the quality of the SED fits is assessed visually. The model parameters cannot be rigorously constrained, and we therefore only identify a parameter set that can represent the SEDs. 

As presented in Table \ref{table_SED}, a hard proton spectrum with an index of $p \sim 1.6 - 1.7$, along with cutoff energies ranging from 30 TeV to 90 TeV, is required to account for the TeV excess observed in the SEDs of the five BL Lacs. As illustrated in Figure \ref{Fig: SED}, this TeV excess is attributed to the decay of $\pi^0$ mesons produced through p$\gamma$ interactions within the ADAFs. The ADAF emission spectra (magenta solid lines) from thermal particles, as well as the synchrotron emission (red dashed lines) originating from cascaded electrons, are significantly overwhelmed by the leptonic jet emission. Additionally, we estimate the proton number density ($N_{\rm p}$) in the energy range [$E_{\rm p,cut}\times 10^{-3}, E_{\rm p,cut}\times 10$] and obtain $N_{\rm p}\sim10^{4-7}$ cm$^{-3}$, as listed in Table \ref{table_SED}. These values do not exceed the typical proton number densities of ADAFs, which range from approximately $10^4$ to $10^{10}$ $\rm cm^{-3}$ from the outer to the inner regions (e.g., \citealt{2014ARA&A..52..529Y, 2015ApJ...806..159K, 2021ApJ...922L..15K, 2021NatCo..12.5615K, 2022ApJ...938...79B}). 

\section{Discussion} \label{sec_dis}

\subsection{Proton Acceleration within ADAF} 

The main challenge for the model is to determine whether protons can be accelerated to such high energies within ADAFs. The maximum attainable energy of charged particles is determined by the Hillas criterion \citep{1984ARA&A..22..425H}, which requires that the size of the acceleration region exceeds the Larmor radius of the particles. According to this criterion, the maximum proton energy can be given by 
\begin{equation}
  E_{\max}=10^{21}\ \mathrm{eV}\left(\frac{R_{\rm L}}{\mathrm{1\ pc}}\right)\left(\frac{B}{ \mathrm{1\ G}}\right)\left(\frac{u}{c}\right),
\end{equation}
where the Larmor radius is assumed to be $R_{\rm L}=R_{\rm o}=10^4R_{\rm S}$, and the magnetic field strength is taken as $B=B_{\rm ADAF} = 1$ G. In the context of diffusive shock acceleration, $u$ denotes the shock velocity with a typical value of $\sim0.1c$ \citep{2007Ap&SS.309..119R}; in magnetic reconnection scenarios, $u$ refers to some flow velocity comparable to the Alfvén speed, which is similarly $\sim0.1c$ \citep{2020NewAR..8901543M}. Therefore, protons can potentially be accelerated to energies of up to $E_{\max} \sim 10^{20}$ eV within ADAFs. 

In RIAFs, the plasma becomes collisionless under conditions of low accretion rates, thereby enabling the presence of non-thermal particles \citep{1994ApJ...428L..13N, 1997ApJ...477..585M, 2012ApJ...754..148T}. In early hadronic models of AGNs, accretion shocks were proposed as potential sites for particle acceleration (e.g., \citealt{1991PhRvL..66.2697S}). However, numerical simulations have not yet confirmed the existence of such shocks \citep{2019PhRvD.100h3014K}. Shocks may arise from the collision of fallback material \citep{2004PhRvD..70l3001A} or from failed winds originating from the accretion disk. In such scenarios, a spectral index of $p \sim 2$ is expected and has been assumed \citep{2004PhRvD..70l3001A, 2020ApJ...891L..33I}. Additionally, the magnetorotational instability (MRI) within RIAFs has long been considered as a mechanism for generating effective viscosity and strong magnetic turbulence, which are essential for angular momentum transport (e.g., \citealt{1991ApJ...376..214B, 2004ApJ...605..321S}). MRI-driven magnetic reconnection can accelerate protons to energies exceeding the PeV range, with spectra as hard as $p < 2$, particularly under conditions of high magnetization and enhanced turbulence near SMBHs (e.g., \citealt{2013ApJ...773..118H, 2019APS..DPPTM9002C, 2021MNRAS.506.1128S}). Alternatively, MRI can generate strong MHD turbulence that scatters protons within RIAFs. This stochastic acceleration mechanism is expected to yield a hard proton spectrum with $p < 1$, with a cutoff occurring around 10--100 PeV due to photo-hadronic interactions (e.g., \citealt{2006ApJ...647..539B, 2008ApJ...681.1725S, 2015ApJ...806..159K, 2021NatCo..12.5615K}). In the case of BL Lacs, the jet is launched from the polar region of the SMBHs, and high-energy protons may also be accelerated via shear acceleration at the jet base \citep{2021MNRAS.506.1128S, 2022ApJ...941L..17M}.

Previous studies have commonly assumed a power-law proton distribution with a spectral index of $ p \sim 2 - 4 $, which is theoretically expected when protons are accelerated via the first-order Fermi mechanism (e.g., \citealt{2019ApJ...880...40I, 2021ApJ...906...51X, 2022ApJ...925L..19C, 2023PhRvD.107j3019X}). In our model, protons must be accelerated to energies ranging from 30 TeV to 90 TeV, with a relatively hard spectral index of $ p \sim 1.6 - 1.7 $. This suggests that magnetic reconnection and/or stochastic acceleration may be the dominant acceleration mechanisms for protons in ADAFs. Based on the stochastic particle acceleration mechanism and a power-law spectrum with $ p < 5/3 $, \citet{2015ApJ...806..159K} demonstrated that nonthermal protons accelerated within the RIAFs of low-luminosity AGNs can generate detectable GeV $\gamma$-rays and neutrons through $pp$ and/or $p\gamma$ interactions.

\subsection{Contributions of the $pp$ Process}

The high-energy protons can also produce TeV photons via the $pp$ process. Therefore, we calculate the $pp$ process using the method developed by \citet{2014PhRvD..90l3014K}, which provides analytical parametrizations for the energy spectra and production rates of $\gamma$-rays resulting from $pp$ interactions. Firstly, we adopt the same proton spectrum and spectral parameter values as those used in the $p\gamma$ process, under the assumption that sufficient cold target protons are present, to calculate the possible $\gamma$-ray spectrum for these BL Lacs. As shown in Figure \ref{Fig: SED}, the $\gamma$-ray spectrum (represented by gray dashed lines) generated by the $pp$ process is insufficiently hard in the TeV energy range to adequately fit the observed data. If the TeV excess is assumed to be predominantly attributed to the $pp$ process, a significantly harder proton spectrum with a higher cutoff energy (exceeding 1 PeV) would be required to accurately reproduce the observational data, compared to the $p\gamma$ process. This scenario imposes more stringent requirements on the underlying particle acceleration mechanism. Therefore, we propose that the $p\gamma$ process is more favorable than the $pp$ process for explaining the TeV excess observed in these BL Lacs.

\subsection{$\gamma\gamma$ Absorption and Cascaded Electron Emission in ADAF}

\begin{figure*}
   \centering
   \includegraphics[angle=0,width=0.6\textwidth]{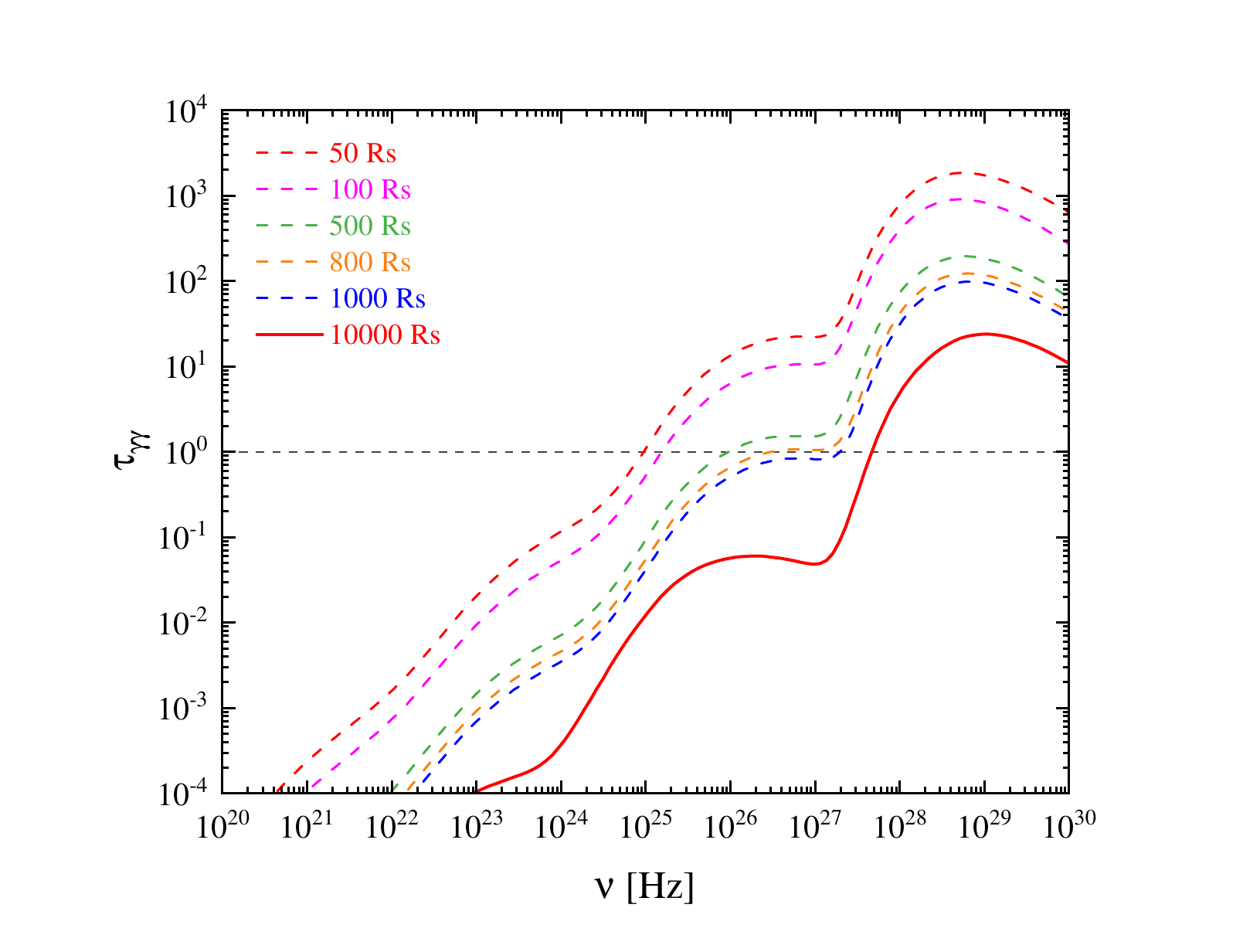}
   \caption{The $\gamma\gamma$ optical depth curves for the ADAF model with varying radii in the source 1ES 0347--121.}
   \label{Fig:tau}
\end{figure*}

High-energy $\gamma$-ray photons originating from $\pi^0$ decay interact with low-energy target photons from the ADAF through the two-photon annihilation process, $\gamma\gamma \rightarrow e^+e^-$. The typical energy of a target photon involved in such an interaction with a TeV photon is given by $\varepsilon_{\rm t}=m_e^2 c^4 /\varepsilon_\gamma \simeq 0.26 \left(1~ \mathrm{TeV} / \varepsilon_\gamma\right)~ \mathrm{eV}$. Using the source 1ES 0347–121 as a case study, we compute the optical depth curves as a function of $\gamma$-ray photon energy for various ADAF sizes, as shown in Figure \ref{Fig:tau}. The results indicate that TeV photons are unable to escape the ADAF due to significant $\gamma\gamma$ absorption when the ADAF radius is too small. A minimum radius of $R_{\rm o}>1000R_{\rm s}$ is required to account for the observed TeV excess. Similar findings have been reported in earlier studies (e.g. \citealt{2022ApJ...939...43E, 2022ApJ...941L..17M, 2023ApJ...954L..49A}). Recently, \citet{2025NatAs.tmp..129L} suggested that the extended corona ($>10^6$ gravitational radii) in radio-quite Seyfert galaxies can produce $\gamma$-rays with energies exceeding several GeV.

The cascaded electrons generated by the $\gamma\gamma$ annihilation process, along with those produced via the p$\gamma$ interaction, can trigger electromagnetic emission, leading to a keV--MeV emission component in the SEDs (e.g., \citealt{2022ApJ...925L..19C, 2022ApJ...941L..17M}). However, due to the weak $\gamma\gamma$ absorption under an ADAF radius of $10^4R_{\rm S}$, the emission from cascaded electrons is found to be insignificant compared to jet radiation in the five BL Lacs, as shown in Figure \ref{Fig: SED}. Therefore, BL Lacs with smaller ADAF radii may serve as promising candidates for testing our model, as these sources do not exhibit a significant TeV excess, yet display a distinct keV--MeV excess in their SEDs.

\subsection{Implications for Variability and Neutrino Detection}

To avoid the $\gamma\gamma$ absorption effect, a large ADAF radius is required in the model to reproduce the observed TeV excess, i.e., $R_{\rm o}\sim10^4R_{\rm s}$, with a minimum of $R_{\rm o}\sim10^3R_{\rm s}$. Within the framework of this model, the derived minimum variability time-scale of TeV emission ranges from several months to years. Therefore, the model is unable to account for the TeV spectra observed during short-term flaring episodes. Among the five SEDs selected in our sample, only the VHE flux from H 2356--309 exhibited low-amplitude variations on timescales of months to years during its TeV detections; however, no significant variability on shorter timescales was observed \citep{2010A&A...516A..56H}. For all other sources, no notable variation in the TeV flux was observed across any timescale \citep{2007A&A...470..475A, 2007A&A...475L...9A, 2007A&A...473L..25A, 2012A&A...538A.103H}. This constitutes the primary rationale for excluding Mrk 421 from our sample: although its SED exhibits a TeV excess component during high-flux states and it is included in the study by \citet{2022ApJ...925L..19C}, its highly variable TeV emission is inconsistent with the assumed emission scenario of the model.

As shown in Formula \ref{eq. neutino}, the model predicts neutrino radiation.Within this theoretical framework, we compute the neutrino flux produced via the $p\gamma$ process and present the results for each source in Figure \ref{Fig: SED}. The total neutrino flux corresponds to the sum of the muon neutrino and muon antineutrino components. The differential sensitivity curves of IceCube for three different declinations, which are obtained from \citet{2019EPJC...79..234A}, are also present in Figure \ref{Fig: SED}. It is evident that, with the exception of 1ES 0229+200, where the predicted neutrino flux slightly overlaps with IceCube's sensitivity curves, the model-predicted fluxes for the remaining four sources fall significantly below IceCube's detection thresholds. We further calculate the neutrino event rate using the formula from \citet{2020PhRvL.124e1103A}:
\begin{equation}
  \frac{d N_{\nu+\bar{\nu}}}{d t}=\int_{0}^{\infty} A_{\rm eff}^{\nu+\bar{\nu}}\left(E_\nu, \delta_{\rm decl}\right) \times \phi_{\nu+\bar{\nu}}\left(E_\nu\right) \mathrm{d} E_\nu,
\end{equation}
where $A_{\rm eff}$ denotes the IceCube point-like source effective area for muon neutrinos and muon antineutrinos at declination $\delta_{\rm decl}$, and $E_{\nu}$ and $\phi_{\nu+\bar{\nu}}\left(E_\nu\right)$ represent the neutrino energy and the differential neutrino energy flux, respectively. The estimated astrophysical neutrino event rates are $2.2$ $\rm yr^{-1}$ for 1ES 0229+200, $0$ $\rm yr^{-1}$ for H2356--309, $2.4\times 10^{-2}$ $\rm yr^{-1}$ for 1ES 1101--232, $1.6\times 10^{-2}$ $\rm yr^{-1}$ for 1ES 0347--121, and $4.7\times 10^{-1}$ $\rm yr^{-1}$ for 1ES 0414+009.

\section{Summary} \label{sec_Summary}

In this work, we explored the potential hadronic origin of the observed TeV excess---defined within the framework of the one-zone SSC model---in the SEDs of five BL Lacs (1ES 0229+200, H2356--309, 1ES 1101--232, 1ES 0347--121, and 1ES 0414+009) using a two-zone lepto-hadronic model. The synchrotron radiation and SSC processes of relativistic electrons within jets account for the emission across the radio, optical, X-ray, and GeV $\gamma$-ray bands. The TeV $\gamma$-ray emission, which exceeds the predictions of the leptonic jet model, is attributed to proton-related emission within ADAFs. Although the $p\gamma$ process is primarily considered, the contribution from $pp$ interactions cannot be entirely ruled out. The model successfully reproduces the broadband SEDs of the five BL Lacs by invoking a population of protons with a hard spectral index of approximately 1.6--1.7 and a cut-off energy in the range of 30--90 TeV within the ADAFs. We propose that these protons are accelerated via magnetic reconnection and/or stochastic acceleration mechanisms within the ADAFs. We also considered the contribution from cascaded electrons generated by hadronic processes; however, this component is significantly overshadowed by the dominant jet emission. Furthermore, for the TeV photons to escape the emission region without being absorbed via $\gamma\gamma$ annihilation process, a large ADAF radius is required, resulting in a weak cascaded emission component in the keV--MeV band. Based on the fitting results and theoretical calculations, we suggest that BL Lacs with smaller ADAF radii may exhibit a prominent keV--MeV spectral bump without a corresponding TeV excess, making them ideal candidates for testing our model. Additionally, the requirement of a large ADAF radius implies that the model cannot readily account for the TeV emission observed during short-timescale flaring events.

\acknowledgments
We thank the anonymous referee for the valuable suggestions and comments. We are grateful to Liang Chen and Da-Bin Lin for the valuable discussion. This work is supported by the National Key R\&D Program of China (grant 2023YFE0117200) and the National Natural Science Foundation of China (grants 12203024, 12022305, 11973050, 12473042, and 12373109).

\clearpage
\bibliography{biblist}{}

@ARTICLE{1995PASP..107..803U,
       author = {{Urry}, C. Megan and {Padovani}, Paolo},
        title = "{Unified Schemes for Radio-Loud Active Galactic Nuclei}",
      journal = {\pasp},
         year = 1995,
        month = sep,
       volume = {107},
        pages = {803},
          doi = {10.1086/133630},
archivePrefix = {arXiv},
       eprint = {astro-ph/9506063},
 primaryClass = {astro-ph},
       adsurl = {https://ui.adsabs.harvard.edu/abs/1995PASP..107..803U}
}

@ARTICLE{1997ApJ...477..585M,
       author = {{Mahadevan}, Rohan},
        title = "{Scaling Laws for Advection-dominated Flows: Applications to Low-Luminosity Galactic Nuclei}",
      journal = {\apj},
         year = 1997,
        month = mar,
       volume = {477},
       number = {2},
        pages = {585-601},
          doi = {10.1086/303727},
archivePrefix = {arXiv},
       eprint = {astro-ph/9609107},
 primaryClass = {astro-ph},
       adsurl = {https://ui.adsabs.harvard.edu/abs/1997ApJ...477..585M}
}

@ARTICLE{2012ApJ...752..157Z,
       author = {{Zhang}, Jin and {Liang}, En-Wei and {Zhang}, Shuang-Nan and {Bai}, J.~M.},
        title = "{Radiation Mechanisms and Physical Properties of GeV-TeV BL Lac Objects}",
      journal = {\apj},
         year = 2012,
        month = jun,
       volume = {752},
       number = {2},
          eid = {157},
        pages = {157},
          doi = {10.1088/0004-637X/752/2/157},
archivePrefix = {arXiv},
       eprint = {1108.0607},
 primaryClass = {astro-ph.HE},
       adsurl = {https://ui.adsabs.harvard.edu/abs/2012ApJ...752..157Z}
}

@ARTICLE{2022ApJ...941...33F,
       author = {{Finke}, Justin D. and {Ajello}, Marco and {Dom{\'\i}nguez}, Alberto and {Desai}, Abhishek and {Hartmann}, Dieter H. and {Paliya}, Vaidehi S. and {Saldana-Lopez}, Alberto},
        title = "{Modeling the Extragalactic Background Light and the Cosmic Star Formation History}",
      journal = {\apj},
         year = 2022,
        month = dec,
       volume = {941},
       number = {1},
          eid = {33},
        pages = {33},
          doi = {10.3847/1538-4357/ac9843},
archivePrefix = {arXiv},
       eprint = {2210.01157},
 primaryClass = {astro-ph.GA},
       adsurl = {https://ui.adsabs.harvard.edu/abs/2022ApJ...941...33F}
}

@ARTICLE{2014MNRAS.439.2933Y,
       author = {{Yan}, Dahai and {Zeng}, Houdun and {Zhang}, Li},
        title = "{The physical properties of Fermi BL Lac objects jets}",
      journal = {\mnras},
         year = 2014,
        month = apr,
       volume = {439},
       number = {3},
        pages = {2933-2942},
          doi = {10.1093/mnras/stu146},
archivePrefix = {arXiv},
       eprint = {1401.5552},
 primaryClass = {astro-ph.HE},
       adsurl = {https://ui.adsabs.harvard.edu/abs/2014MNRAS.439.2933Y}
}

@ARTICLE{2017ApJ...842..129C,
       author = {{Chen}, Liang},
        title = "{On the Origin of the Hard X-Ray Excess of High-Synchrotron-Peaked BL Lac Object Mrk 421}",
      journal = {\apj},
         year = 2017,
        month = jun,
       volume = {842},
       number = {2},
          eid = {129},
        pages = {129},
          doi = {10.3847/1538-4357/aa7744},
archivePrefix = {arXiv},
       eprint = {1706.04611},
 primaryClass = {astro-ph.HE},
       adsurl = {https://ui.adsabs.harvard.edu/abs/2017ApJ...842..129C}
}

@ARTICLE{2014Natur.515..376G,
       author = {{Ghisellini}, G. and {Tavecchio}, F. and {Maraschi}, L. and {Celotti}, A. and {Sbarrato}, T.},
        title = "{The power of relativistic jets is larger than the luminosity of their accretion disks}",
      journal = {\nat},
         year = 2014,
        month = nov,
       volume = {515},
       number = {7527},
        pages = {376-378},
          doi = {10.1038/nature13856},
archivePrefix = {arXiv},
       eprint = {1411.5368},
 primaryClass = {astro-ph.HE},
       adsurl = {https://ui.adsabs.harvard.edu/abs/2014Natur.515..376G}
}

@ARTICLE{2022ApJ...925...97P,
       author = {{Pei}, Zhiyuan and {Fan}, Junhui and {Yang}, Jianghe and {Huang}, Danyi and {Li}, Ziyan},
        title = "{The Estimation of Fundamental Physics Parameters for Fermi-LAT Blazars}",
      journal = {\apj},
         year = 2022,
        month = jan,
       volume = {925},
       number = {1},
          eid = {97},
        pages = {97},
          doi = {10.3847/1538-4357/ac3aeb},
archivePrefix = {arXiv},
       eprint = {2112.00530},
 primaryClass = {astro-ph.HE},
       adsurl = {https://ui.adsabs.harvard.edu/abs/2022ApJ...925...97P}
}

@ARTICLE{2005ApJ...630..130B,
       author = {{B{\l}a{\.z}ejowski}, M. and {Blaylock}, G. and {Bond}, I.~H. and {Bradbury}, S.~M. and {Buckley}, J.~H. and {Carter-Lewis}, D.~A. and {Celik}, O. and {Cogan}, P. and {Cui}, W. and {Daniel}, M. and {Duke}, C. and {Falcone}, A. and {Fegan}, D.~J. and {Fegan}, S.~J. and {Finley}, J.~P. and {Fortson}, L. and {Gammell}, S. and {Gibbs}, K. and {Gillanders}, G.~G. and {Grube}, J. and {Gutierrez}, K. and {Hall}, J. and {Hanna}, D. and {Holder}, J. and {Horan}, D. and {Humensky}, B. and {Kenny}, G. and {Kertzman}, M. and {Kieda}, D. and {Kildea}, J. and {Knapp}, J. and {Kosack}, K. and {Krawczynski}, H. and {Krennrich}, F. and {Lang}, M. and {LeBohec}, S. and {Linton}, E. and {Lloyd-Evans}, J. and {Maier}, G. and {Mendoza}, D. and {Milovanovic}, A. and {Moriarty}, P. and {Nagai}, T.~N. and {Ong}, R.~A. and {Power-Mooney}, B. and {Quinn}, J. and {Quinn}, M. and {Ragan}, K. and {Reynolds}, P.~T. and {Rebillot}, P. and {Rose}, H.~J. and {Schroedter}, M. and {Sembroski}, G.~H. and {Swordy}, S.~P. and {Syson}, A. and {Valcarel}, L. and {Vassiliev}, V.~V. and {Wakely}, S.~P. and {Walker}, G. and {Weekes}, T.~C. and {White}, R. and {Zweerink}, J. and {VERITAS Collaboration} and {Mochejska}, B. and {Smith}, B. and {Aller}, M. and {Aller}, H. and {Ter{\"a}sranta}, H. and {Boltwood}, P. and {Sadun}, A. and {Stanek}, K. and {Adams}, E. and {Foster}, J. and {Hartman}, J. and {Lai}, K. and {B{\"o}ttcher}, M. and {Reimer}, A. and {Jung}, I.},
        title = "{A Multiwavelength View of the TeV Blazar Markarian 421: Correlated Variability, Flaring, and Spectral Evolution}",
      journal = {\apj},
         year = 2005,
        month = sep,
       volume = {630},
       number = {1},
        pages = {130-141},
          doi = {10.1086/431925},
archivePrefix = {arXiv},
       eprint = {astro-ph/0505325},
 primaryClass = {astro-ph},
       adsurl = {https://ui.adsabs.harvard.edu/abs/2005ApJ...630..130B}
}

@ARTICLE{2010MNRAS.401.1570T,
       author = {{Tavecchio}, F. and {Ghisellini}, G. and {Ghirlanda}, G. and {Foschini}, L. and {Maraschi}, L.},
        title = "{TeV BL Lac objects at the dawn of the Fermi era}",
      journal = {\mnras},
         year = 2010,
        month = jan,
       volume = {401},
       number = {3},
        pages = {1570-1586},
          doi = {10.1111/j.1365-2966.2009.15784.x},
archivePrefix = {arXiv},
       eprint = {0909.0651},
 primaryClass = {astro-ph.HE},
       adsurl = {https://ui.adsabs.harvard.edu/abs/2010MNRAS.401.1570T}
}

@ARTICLE{2019ARA&A..57..467B,
       author = {{Blandford}, Roger and {Meier}, David and {Readhead}, Anthony},
        title = "{Relativistic Jets from Active Galactic Nuclei}",
      journal = {\araa},
         year = 2019,
        month = aug,
       volume = {57},
        pages = {467-509},
          doi = {10.1146/annurev-astro-081817-051948},
archivePrefix = {arXiv},
       eprint = {1812.06025},
 primaryClass = {astro-ph.HE},
       adsurl = {https://ui.adsabs.harvard.edu/abs/2019ARA&A..57..467B}
}

@ARTICLE{2002ApJ...579..554W,
       author = {{Wang}, Jian-Min and {Staubert}, R{\"u}diger and {Ho}, Luis C.},
        title = "{The Accretion Rates and Spectral Energy Distributions of BL Lacertae Objects}",
      journal = {\apj},
         year = 2002,
        month = nov,
       volume = {579},
       number = {2},
        pages = {554-559},
          doi = {10.1086/342875},
archivePrefix = {arXiv},
       eprint = {astro-ph/0207305},
 primaryClass = {astro-ph},
       adsurl = {https://ui.adsabs.harvard.edu/abs/2002ApJ...579..554W}
}

@INPROCEEDINGS{1998tbha.conf..148N,
       author = {{Narayan}, R. and {Mahadevan}, R. and {Quataert}, E.},
        title = "{Advection-dominated accretion around black holes}",
    booktitle = {Theory of Black Hole Accretion Disks},
         year = 1998,
       editor = {{Abramowicz}, M.~A. and {Bj{\"o}rnsson}, G. and {Pringle}, J.~E.},
        month = jan,
        pages = {148-182},
          doi = {10.48550/arXiv.astro-ph/9803141},
archivePrefix = {arXiv},
       eprint = {astro-ph/9803141},
 primaryClass = {astro-ph},
       adsurl = {https://ui.adsabs.harvard.edu/abs/1998tbha.conf..148N}
}

@ARTICLE{2024ApJ...967..104Z,
       author = {{Zhao}, X.~Z. and {Yang}, H.~Y. and {Zheng}, Y.~G. and {Kang}, S.~J.},
        title = "{The Energy Budget in the Jet of High-frequency Peaked BL Lacertae Objects}",
      journal = {\apj},
         year = 2024,
        month = jun,
       volume = {967},
       number = {2},
          eid = {104},
        pages = {104},
          doi = {10.3847/1538-4357/ad3ba9},
archivePrefix = {arXiv},
       eprint = {2406.01046},
 primaryClass = {astro-ph.HE},
       adsurl = {https://ui.adsabs.harvard.edu/abs/2024ApJ...967..104Z}
}

@ARTICLE{2023MNRAS.526.4079C,
       author = {{Chen}, Yongyun and {Gu}, Qiusheng and {Fan}, Junhui and {Yu}, Xiaoling and {Ding}, Nan and {Guo}, Xiaotong and {Xiong}, Dingrong},
        title = "{Jet power extracted from ADAFs and the application to Fermi BL Lacertae objects}",
      journal = {\mnras},
         year = 2023,
        month = dec,
       volume = {526},
       number = {3},
        pages = {4079-4092},
          doi = {10.1093/mnras/stad2623},
archivePrefix = {arXiv},
       eprint = {2308.15707},
 primaryClass = {astro-ph.HE},
       adsurl = {https://ui.adsabs.harvard.edu/abs/2023MNRAS.526.4079C}
}

@ARTICLE{1984ARA&A..22..425H,
       author = {{Hillas}, A.~M.},
        title = "{The Origin of Ultra-High-Energy Cosmic Rays}",
      journal = {\araa},
         year = 1984,
        month = jan,
       volume = {22},
        pages = {425-444},
          doi = {10.1146/annurev.aa.22.090184.002233},
       adsurl = {https://ui.adsabs.harvard.edu/abs/1984ARA&A..22..425H}
}

@ARTICLE{2022ApJ...938...79B,
       author = {{Boughelilba}, Margot and {Reimer}, Anita and {Merten}, Lukas},
        title = "{Lepto-hadronic Jet-disk Model for the Multiwavelength SED of M87}",
      journal = {\apj},
         year = 2022,
        month = oct,
       volume = {938},
       number = {1},
          eid = {79},
        pages = {79},
          doi = {10.3847/1538-4357/ac8e64},
archivePrefix = {arXiv},
       eprint = {2208.14756},
 primaryClass = {astro-ph.HE},
       adsurl = {https://ui.adsabs.harvard.edu/abs/2022ApJ...938...79B}
}

@ARTICLE{1982Natur.295...17R,
       author = {{Rees}, M.~J. and {Begelman}, M.~C. and {Blandford}, R.~D. and {Phinney}, E.~S.},
        title = "{Ion-supported tori and the origin of radio jets}",
      journal = {\nat},
         year = 1982,
        month = jan,
       volume = {295},
       number = {5844},
        pages = {17-21},
          doi = {10.1038/295017a0},
       adsurl = {https://ui.adsabs.harvard.edu/abs/1982Natur.295...17R}
}

@ARTICLE{1977ApJ...214..840I,
       author = {{Ichimaru}, S.},
        title = "{Bimodal behavior of accretion disks: theory and application to Cygnus X-1 transitions.}",
      journal = {\apj},
         year = 1977,
        month = jun,
       volume = {214},
        pages = {840-855},
          doi = {10.1086/155314},
       adsurl = {https://ui.adsabs.harvard.edu/abs/1977ApJ...214..840I}
}

@ARTICLE{2005ApJ...620..905Y,
       author = {{Yuan}, Feng and {Cui}, Wei and {Narayan}, Ramesh},
        title = "{An Accretion-Jet Model for Black Hole Binaries: Interpreting the Spectral and Timing Features of XTE J1118+480}",
      journal = {\apj},
         year = 2005,
        month = feb,
       volume = {620},
       number = {2},
        pages = {905-914},
          doi = {10.1086/427206},
archivePrefix = {arXiv},
       eprint = {astro-ph/0407612},
 primaryClass = {astro-ph},
       adsurl = {https://ui.adsabs.harvard.edu/abs/2005ApJ...620..905Y}
}

@ARTICLE{2007ApJ...659..541Y,
       author = {{Yuan}, Feng and {Zdziarski}, Andrzej A. and {Xue}, Yongquan and {Wu}, Xue-Bing},
        title = "{Modeling the Hard States of XTE J1550-564 during Its 2000 Outburst}",
      journal = {\apj},
         year = 2007,
        month = apr,
       volume = {659},
       number = {1},
        pages = {541-548},
          doi = {10.1086/512078},
archivePrefix = {arXiv},
       eprint = {astro-ph/0608552},
 primaryClass = {astro-ph},
       adsurl = {https://ui.adsabs.harvard.edu/abs/2007ApJ...659..541Y}
}

@ARTICLE{2014MNRAS.438.2804N,
       author = {{Nemmen}, Rodrigo S. and {Storchi-Bergmann}, Thaisa and {Eracleous}, Michael},
        title = "{Spectral models for low-luminosity active galactic nuclei in LINERs: the role of advection-dominated accretion and jets}",
      journal = {\mnras},
         year = 2014,
        month = mar,
       volume = {438},
       number = {4},
        pages = {2804-2827},
          doi = {10.1093/mnras/stt2388},
archivePrefix = {arXiv},
       eprint = {1312.1982},
 primaryClass = {astro-ph.HE},
       adsurl = {https://ui.adsabs.harvard.edu/abs/2014MNRAS.438.2804N}
}

@ARTICLE{2021NatCo..12.5615K,
       author = {{Kimura}, Shigeo S. and {Murase}, Kohta and {M{\'e}sz{\'a}ros}, P{\'e}ter},
        title = "{Soft gamma rays from low accreting supermassive black holes and connection to energetic neutrinos}",
      journal = {Nature Communications},
         year = 2021,
        month = sep,
       volume = {12},
          eid = {5615},
        pages = {5615},
          doi = {10.1038/s41467-021-25111-7},
archivePrefix = {arXiv},
       eprint = {2005.01934},
 primaryClass = {astro-ph.HE},
       adsurl = {https://ui.adsabs.harvard.edu/abs/2021NatCo..12.5615K}
}

@ARTICLE{2019PhRvD.100h3014K,
       author = {{Kimura}, Shigeo S. and {Murase}, Kohta and {M{\'e}sz{\'a}ros}, Peter},
        title = "{Multimessenger tests of cosmic-ray acceleration in radiatively inefficient accretion flows}",
      journal = {\prd},
         year = 2019,
        month = oct,
       volume = {100},
       number = {8},
          eid = {083014},
        pages = {083014},
          doi = {10.1103/PhysRevD.100.083014},
archivePrefix = {arXiv},
       eprint = {1908.08421},
 primaryClass = {astro-ph.HE},
       adsurl = {https://ui.adsabs.harvard.edu/abs/2019PhRvD.100h3014K}
}

@ARTICLE{2015ApJ...806..159K,
       author = {{Kimura}, Shigeo S. and {Murase}, Kohta and {Toma}, Kenji},
        title = "{Neutrino and Cosmic-Ray Emission and Cumulative Background from Radiatively Inefficient Accretion Flows in Low-luminosity Active Galactic Nuclei}",
      journal = {\apj},
         year = 2015,
        month = jun,
       volume = {806},
       number = {2},
          eid = {159},
        pages = {159},
          doi = {10.1088/0004-637X/806/2/159},
archivePrefix = {arXiv},
       eprint = {1411.3588},
 primaryClass = {astro-ph.HE},
       adsurl = {https://ui.adsabs.harvard.edu/abs/2015ApJ...806..159K}
}

@ARTICLE{2003ApJ...598..301Y,
       author = {{Yuan}, Feng and {Quataert}, Eliot and {Narayan}, Ramesh},
        title = "{Nonthermal Electrons in Radiatively Inefficient Accretion Flow Models of Sagittarius A*}",
      journal = {\apj},
         year = 2003,
        month = nov,
       volume = {598},
       number = {1},
        pages = {301-312},
          doi = {10.1086/378716},
archivePrefix = {arXiv},
       eprint = {astro-ph/0304125},
 primaryClass = {astro-ph},
       adsurl = {https://ui.adsabs.harvard.edu/abs/2003ApJ...598..301Y}
}

@ARTICLE{2014ARA&A..52..529Y,
       author = {{Yuan}, Feng and {Narayan}, Ramesh},
        title = "{Hot Accretion Flows Around Black Holes}",
      journal = {\araa},
         year = 2014,
        month = aug,
       volume = {52},
        pages = {529-588},
          doi = {10.1146/annurev-astro-082812-141003},
archivePrefix = {arXiv},
       eprint = {1401.0586},
 primaryClass = {astro-ph.HE},
       adsurl = {https://ui.adsabs.harvard.edu/abs/2014ARA&A..52..529Y}
}

@ARTICLE{2003ApJ...585L..23F,
       author = {{Fan}, J.~H.},
        title = "{Relation between BL Lacertae Objects and Flat-Spectrum Radio Quasars}",
      journal = {\apjl},
         year = 2003,
        month = mar,
       volume = {585},
       number = {1},
        pages = {L23-L24},
          doi = {10.1086/374033},
       adsurl = {https://ui.adsabs.harvard.edu/abs/2003ApJ...585L..23F}
}

@ARTICLE{1991ApJ...374..431S,
       author = {{Stickel}, M. and {Padovani}, P. and {Urry}, C.~M. and {Fried}, J.~W. and {Kuehr}, H.},
        title = "{The Complete Sample of 1 Jansky BL Lacertae Objects. I. Summary Properties}",
      journal = {\apj},
         year = 1991,
        month = jun,
       volume = {374},
        pages = {431},
          doi = {10.1086/170133},
       adsurl = {https://ui.adsabs.harvard.edu/abs/1991ApJ...374..431S}
}

@ARTICLE{1992ApJ...397L...5M,
       author = {{Maraschi}, L. and {Ghisellini}, G. and {Celotti}, A.},
        title = "{A Jet Model for the Gamma-Ray--emitting Blazar 3C 279}",
      journal = {\apjl},
         year = 1992,
        month = sep,
       volume = {397},
        pages = {L5},
          doi = {10.1086/186531},
       adsurl = {https://ui.adsabs.harvard.edu/abs/1992ApJ...397L...5M}
}

@ARTICLE{2022ApJ...925L..19C,
       author = {{Cheng}, Ji-Gui and {Huang}, Xiao-Li and {Wang}, Ze-Rui and {Huang}, Jian-Kun and {Liang}, En-Wei},
        title = "{TeV and keV-MeV Excesses as Probes for Hadronic Process in BL Lacertaes}",
      journal = {\apjl},
         year = 2022,
        month = feb,
       volume = {925},
       number = {2},
          eid = {L19},
        pages = {L19},
          doi = {10.3847/2041-8213/ac4d8e},
archivePrefix = {arXiv},
       eprint = {2201.08148},
 primaryClass = {astro-ph.HE},
       adsurl = {https://ui.adsabs.harvard.edu/abs/2022ApJ...925L..19C}
}

@ARTICLE{2020ApJS..247...16A,
       author = {{Acciari}, V.~A. and {Ansoldi}, S. and {Antonelli}, L.~A. and {Engels}, A. Arbet and {Asano}, K. and {Baack}, D. and {Babi{\'c}}, A. and {Banerjee}, B. and {Barres de Almeida}, U. and {Barrio}, J.~A. and {Becerra Gonz{\'a}lez}, J. and {Bednarek}, W. and {Bellizzi}, L. and {Bernardini}, E. and {Berti}, A. and {Besenrieder}, J. and {Bhattacharyya}, W. and {Bigongiari}, C. and {Biland}, A. and {Blanch}, O. and {Bonnoli}, G. and {Bo{\v{s}}njak}, {\v{Z}}. and {Busetto}, G. and {Carosi}, R. and {Ceribella}, G. and {Cerruti}, M. and {Chai}, Y. and {Chilingaryan}, A. and {Cikota}, S. and {Colak}, S.~M. and {Colin}, U. and {Colombo}, E. and {Contreras}, J.~L. and {Cortina}, J. and {Covino}, S. and {D'Elia}, V. and {Da Vela}, P. and {Dazzi}, F. and {De Angelis}, A. and {De Lotto}, B. and {Delfino}, M. and {Delgado}, J. and {Depaoli}, D. and {Di Pierro}, F. and {Di Venere}, L. and {Do Souto Espi{\~n}eira}, E. and {Dominis Prester}, D. and {Donini}, A. and {Dorner}, D. and {Doro}, M. and {Elsaesser}, D. and {Ramazani}, V. Fallah and {Fattorini}, A. and {Ferrara}, G. and {Fidalgo}, D. and {Foffano}, L. and {Fonseca}, M.~V. and {Font}, L. and {Fruck}, C. and {Fukami}, S. and {Garc{\'\i}a L{\'o}pez}, R.~J. and {Garczarczyk}, M. and {Gasparyan}, S. and {Gaug}, M. and {Giglietto}, N. and {Giordano}, F. and {Godinovi{\'c}}, N. and {Green}, D. and {Guberman}, D. and {Hadasch}, D. and {Hahn}, A. and {Herrera}, J. and {Hoang}, J. and {Hrupec}, D. and {H{\"u}tten}, M. and {Inada}, T. and {Inoue}, S. and {Ishio}, K. and {Iwamura}, Y. and {Jouvin}, L. and {Kerszberg}, D. and {Kubo}, H. and {Kushida}, J. and {Lamastra}, A. and {Lelas}, D. and {Leone}, F. and {Lindfors}, E. and {Lombardi}, S. and {Longo}, F. and {L{\'o}pez}, M. and {L{\'o}pez-Coto}, R. and {L{\'o}pez-Oramas}, A. and {Loporchio}, S. and {Machado de Oliveira Fraga}, B. and {Maggio}, C. and {Majumdar}, P. and {Makariev}, M. and {Mallamaci}, M. and {Maneva}, G. and {Manganaro}, M. and {Mannheim}, K. and {Maraschi}, L. and {Mariotti}, M. and {Mart{\'\i}nez}, M. and {Mazin}, D. and {Mi{\'c}anovi{\'c}}, S. and {Miceli}, D. and {Minev}, M. and {Miranda}, J.~M. and {Mirzoyan}, R. and {Molina}, E. and {Moralejo}, A. and {Morcuende}, D. and {Moreno}, V. and {Moretti}, E. and {Munar-Adrover}, P. and {Neustroev}, V. and {Nigro}, C. and {Nilsson}, K. and {Ninci}, D. and {Nishijima}, K. and {Noda}, K. and {Nogu{\'e}s}, L. and {Nozaki}, S. and {Paiano}, S. and {Palatiello}, M. and {Paneque}, D. and {Paoletti}, R. and {Paredes}, J.~M. and {Pe{\~n}il}, P. and {Peresano}, M. and {Persic}, M. and {Prada Moroni}, P.~G. and {Prandini}, E. and {Puljak}, I. and {Rhode}, W. and {Rib{\'o}}, M. and {Rico}, J. and {Righi}, C. and {Rugliancich}, A. and {Saha}, L. and {Sahakyan}, N. and {Saito}, T. and {Sakurai}, S. and {Satalecka}, K. and {Schmidt}, K. and {Schweizer}, T. and {Sitarek}, J. and {{\v{S}}nidari{\'c}}, I. and {Sobczynska}, D. and {Somero}, A. and {Stamerra}, A. and {Strom}, D. and {Strzys}, M. and {Suda}, Y. and {Suri{\'c}}, T. and {Takahashi}, M. and {Tavecchio}, F. and {Temnikov}, P. and {Terzi{\'c}}, T. and {Teshima}, M. and {Torres-Alb{\`a}}, N. and {Tosti}, L. and {Vagelli}, V. and {van Scherpenberg}, J. and {Vanzo}, G. and {Vazquez Acosta}, M. and {Vigorito}, C.~F. and {Vitale}, V. and {Vovk}, I. and {Will}, M. and {Zari{\'c}}, D. and {Arcaro}, C. and {Carosi}, A. and {D'Ammando}, F. and {Tombesi}, F. and {Lohfink}, A.},
        title = "{New Hard-TeV Extreme Blazars Detected with the MAGIC Telescopes}",
      journal = {\apjs},
         year = 2020,
        month = mar,
       volume = {247},
       number = {1},
          eid = {16},
        pages = {16},
          doi = {10.3847/1538-4365/ab5b98},
archivePrefix = {arXiv},
       eprint = {1911.06680},
 primaryClass = {astro-ph.HE},
       adsurl = {https://ui.adsabs.harvard.edu/abs/2020ApJS..247...16A}
}

@ARTICLE{2015MNRAS.448..910C,
       author = {{Cerruti}, M. and {Zech}, A. and {Boisson}, C. and {Inoue}, S.},
        title = "{A hadronic origin for ultra-high-frequency-peaked BL Lac objects}",
      journal = {\mnras},
         year = 2015,
        month = mar,
       volume = {448},
       number = {1},
        pages = {910-927},
          doi = {10.1093/mnras/stu2691},
archivePrefix = {arXiv},
       eprint = {1411.5968},
 primaryClass = {astro-ph.HE},
       adsurl = {https://ui.adsabs.harvard.edu/abs/2015MNRAS.448..910C}
}

@ARTICLE{2013ApJ...768...54B,
       author = {{B{\"o}ttcher}, M. and {Reimer}, A. and {Sweeney}, K. and {Prakash}, A.},
        title = "{Leptonic and Hadronic Modeling of Fermi-detected Blazars}",
      journal = {\apj},
         year = 2013,
        month = may,
       volume = {768},
       number = {1},
          eid = {54},
        pages = {54},
          doi = {10.1088/0004-637X/768/1/54},
archivePrefix = {arXiv},
       eprint = {1304.0605},
 primaryClass = {astro-ph.HE},
       adsurl = {https://ui.adsabs.harvard.edu/abs/2013ApJ...768...54B}
}

@ARTICLE{2021ApJ...906...51X,
       author = {{Xue}, Rui and {Liu}, Ruo-Yu and {Wang}, Ze-Rui and {Ding}, Nan and {Wang}, Xiang-Yu},
        title = "{A Two-zone Blazar Radiation Model for ``Orphan'' Neutrino Flares}",
      journal = {\apj},
         year = 2021,
        month = jan,
       volume = {906},
       number = {1},
          eid = {51},
        pages = {51},
          doi = {10.3847/1538-4357/abc886},
archivePrefix = {arXiv},
       eprint = {2011.03681},
 primaryClass = {astro-ph.HE},
       adsurl = {https://ui.adsabs.harvard.edu/abs/2021ApJ...906...51X}
}

@ARTICLE{2007A&A...475L...9A,
       author = {{Aharonian}, F. and {Akhperjanian}, A.~G. and {Barres de Almeida}, U. and {Bazer-Bachi}, A.~R. and {Behera}, B. and {Beilicke}, M. and {Benbow}, W. and {Bernl{\"o}hr}, K. and {Boisson}, C. and {Bolz}, O. and {Borrel}, V. and {Braun}, I. and {Brion}, E. and {Brown}, A.~M. and {B{\"u}hler}, R. and {Bulik}, T. and {B{\"u}sching}, I. and {Boutelier}, T. and {Carrigan}, S. and {Chadwick}, P.~M. and {Chounet}, L. -M. and {Clapson}, A.~C. and {Coignet}, G. and {Cornils}, R. and {Costamante}, L. and {Dalton}, M. and {Degrange}, B. and {Dickinson}, H.~J. and {Djannati-Ata{\"\i}}, A. and {Domainko}, W. and {O'C. Drury}, L. and {Dubois}, F. and {Dubus}, G. and {Dyks}, J. and {Egberts}, K. and {Emmanoulopoulos}, D. and {Espigat}, P. and {Farnier}, C. and {Feinstein}, F. and {Fiasson}, A. and {F{\"o}rster}, A. and {Fontaine}, G. and {Funk}, Seb. and {F{\"u}{\ss}ling}, M. and {Gallant}, Y.~A. and {Giebels}, B. and {Glicenstein}, J.~F. and {Gl{\"u}ck}, B. and {Goret}, P. and {Hadjichristidis}, C. and {Hauser}, D. and {Hauser}, M. and {Heinzelmann}, G. and {Henri}, G. and {Hermann}, G. and {Hinton}, J.~A. and {Hoffmann}, A. and {Hofmann}, W. and {Holleran}, M. and {Hoppe}, S. and {Horns}, D. and {Jacholkowska}, A. and {de Jager}, O.~C. and {Jung}, I. and {Katarzy{\'n}ski}, K. and {Kendziorra}, E. and {Kerschhaggl}, M. and {Kh{\'e}lifi}, B. and {Keogh}, D. and {Komin}, Nu. and {Kosack}, K. and {Lamanna}, G. and {Latham}, I.~J. and {Lemi{\`e}re}, A. and {Lemoine-Goumard}, M. and {Lenain}, J. -P. and {Lohse}, T. and {Martin}, J.~M. and {Martineau-Huynh}, O. and {Marcowith}, A. and {Masterson}, C. and {Maurin}, D. and {Maurin}, G. and {McComb}, T.~J.~L. and {Moderski}, R. and {Moulin}, E. and {de Naurois}, M. and {Nedbal}, D. and {Nolan}, S.~J. and {Ohm}, S. and {Olive}, J. -P. and {de O{\~n}a Wilhelmi}, E. and {Orford}, K.~J. and {Osborne}, J.~L. and {Ostrowski}, M. and {Panter}, M. and {Pedaletti}, G. and {Pelletier}, G. and {Petrucci}, P. -O. and {Pita}, S. and {P{\"u}hlhofer}, G. and {Punch}, M. and {Ranchon}, S. and {Raubenheimer}, B.~C. and {Raue}, M. and {Rayner}, S.~M. and {Renaud}, M. and {Ripken}, J. and {Rob}, L. and {Rolland}, L. and {Rosier-Lees}, S. and {Rowell}, G. and {Rudak}, B. and {Ruppel}, J. and {Sahakian}, V. and {Santangelo}, A. and {Schlickeiser}, R. and {Sch{\"o}ck}, F. and {Schr{\"o}der}, R. and {Schwanke}, U. and {Schwarzburg}, S. and {Schwemmer}, S. and {Shalchi}, A. and {Sol}, H. and {Spangler}, D. and {Stawarz}, {\L}. and {Steenkamp}, R. and {Stegmann}, C. and {Superina}, G. and {Tam}, P.~H. and {Tavernet}, J. -P. and {Terrier}, R. and {van Eldik}, C. and {Vasileiadis}, G. and {Venter}, C. and {Vialle}, J.~P. and {Vincent}, P. and {Vivier}, M. and {V{\"o}lk}, H.~J. and {Volpe}, F. and {Wagner}, S.~J. and {Ward}, M. and {Zdziarski}, A.~A. and {Zech}, A.},
        title = "{New constraints on the mid-IR EBL from the HESS discovery of VHE {\ensuremath{\gamma}}-rays from 1ES 0229+200}",
      journal = {\aap},
         year = 2007,
        month = nov,
       volume = {475},
       number = {2},
        pages = {L9-L13},
          doi = {10.1051/0004-6361:20078462},
archivePrefix = {arXiv},
       eprint = {0709.4584},
 primaryClass = {astro-ph},
       adsurl = {https://ui.adsabs.harvard.edu/abs/2007A&A...475L...9A}
}

@ARTICLE{2005ApJ...631..762W,
       author = {{Woo}, Jong-Hak and {Urry}, C. Megan and {van der Marel}, Roeland P. and {Lira}, Paulina and {Maza}, Jose},
        title = "{Black Hole Masses and Host Galaxy Evolution of Radio-Loud Active Galactic Nuclei}",
      journal = {\apj},
         year = 2005,
        month = oct,
       volume = {631},
       number = {2},
        pages = {762-772},
          doi = {10.1086/432681},
archivePrefix = {arXiv},
       eprint = {astro-ph/0506316},
 primaryClass = {astro-ph},
       adsurl = {https://ui.adsabs.harvard.edu/abs/2005ApJ...631..762W}
}

@ARTICLE{2012A&A...538A.103H,
       author = {{H.~E.~S.~S. Collaboration} and {Abramowski}, A. and {Acero}, F. and {Aharonian}, F. and {Akhperjanian}, A.~G. and {Anton}, G. and {Balzer}, A. and {Barnacka}, A. and {Barres de Almeida}, U. and {Becherini}, Y. and {Becker}, J. and {Behera}, B. and {Bernloehr}, K. and {Birsin}, E. and {Biteau}, J. and {Bochow}, A. and {Boisson}, C. and {Bolmont}, J. and {Bordas}, P. and {Brucker}, J. and {Brun}, F. and {Brun}, P. and {Bulik}, T. and {Buesching}, I. and {Carrigan}, S. and {Casanova}, S. and {Cerruti}, M. and {Chadwick}, P.~M. and {Charbonnier}, A. and {Chaves}, R.~C.~G. and {Cheesebrough}, A. and {Chounet}, L. -M. and {Clapson}, A.~C. and {Coignet}, G. and {Cologna}, G. and {Conrad}, J. and {Dalton}, M. and {Daniel}, M.~K. and {Davids}, I.~D. and {Degrange}, B. and {Deil}, C. and {Dickinson}, H.~J. and {Djannati-Ataie}, A. and {Domainko}, W. and {Drury}, L. O'c. and {Dubois}, F. and {Dubus}, G. and {Dutson}, K. and {Dyks}, J. and {Dyrda}, M. and {Egberts}, K. and {Eger}, P. and {Espigat}, P. and {Fallon}, L. and {Farnier}, C. and {Feinstein}, F. and {Fernandes}, M.~V. and {Fiasson}, A. and {Fontaine}, G. and {Foerster}, A. and {Fuesling}, M. and {Gallant}, Y.~A. and {Gast}, H. and {Gerard}, L. and {Gerbig}, D. and {Giebels}, B. and {Glicenstein}, J.~F. and {Glueck}, B. and {Goret}, P. and {Goering}, D. and {Haeffner}, S. and {Hague}, J.~D. and {Hampf}, D. and {Hauser}, M. and {Heinz}, S. and {Heinzelmann}, G. and {Henri}, G. and {Hermann}, G. and {Hinton}, J.~A. and {Hoffmann}, A. and {Hofmann}, W. and {Hofverberg}, P. and {Holler}, M. and {Horns}, D. and {Jacholkowska}, A. and {de Jager}, O.~C. and {Jahn}, C. and {Jamrozy}, M. and {Jung}, I. and {Kastendieck}, M.~A. and {Katarzynski}, K. and {Katz}, U. and {Kaufmann}, S. and {Keogh}, D. and {Khangulyan}, D. and {Khelifi}, B. and {Klochkov}, D. and {Kluzniak}, W. and {Kneiske}, T. and {Komin}, Nu. and {Kosack}, K. and {Kossakowski}, R. and {Laffon}, H. and {Lamanna}, G. and {Lennarz}, D. and {Lohse}, T. and {Lopatin}, A. and {Lu}, C. -C. and {Marandon}, V. and {Marcowith}, A. and {Masbou}, J. and {Maurin}, D. and {Maxted}, N. and {Mayer}, M. and {McComb}, T.~J.~L. and {Medina}, M.~C. and {Mehault}, J. and {Moderski}, R. and {Moulin}, E. and {Naumann}, C.~L. and {Naumann-Godo}, M. and {de Naurois}, M. and {Nedbal}, D. and {Nekrassov}, D. and {Nguyen}, N. and {Nicholas}, B. and {Niemiec}, J. and {Nolan}, S.~J. and {Ohm}, S. and {de Ona Wilhelmi}, E. and {Opitz}, B. and {Ostrowski}, M. and {Oya}, I. and {Panter}, M. and {Paz Arribas}, M. and {Pedaletti}, G. and {Pelletier}, G. and {Petrucci}, P. -O. and {Pita}, S. and {Puehlhofer}, G. and {Punch}, M. and {Quirrenbach}, A. and {Raue}, M. and {Rayner}, S.~M. and {Reimer}, A. and {Reimer}, O. and {Renaud}, M. and {de Los Reyes}, R. and {Rieger}, F. and {Ripken}, J. and {Rob}, L. and {Rosier-Lees}, S. and {Rowell}, G. and {Rudak}, B. and {Rulten}, C.~B. and {Ruppel}, J. and {Sahakian}, V. and {Sanchez}, D.~A. and {Santangelo}, A. and {Schlickeiser}, R. and {Schoeck}, F.~M. and {Schulz}, A. and {Schwanke}, U. and {Schwarzburg}, S. and {Schwemmer}, S. and {Sheidaei}, F. and {Sikora}, M. and {Skilton}, J.~L. and {Sol}, H. and {Spengler}, G. and {Stawarz}, L. and {Steenkamp}, R. and {Stegmann}, C. and {Stinzing}, F. and {Stycz}, K. and {Sushch}, I. and {Szostek}, A. and {Tavernet}, J. -P. and {Terrier}, R. and {Tluczykont}, M. and {Valerius}, K. and {van Eldik}, C. and {Vasileiadis}, G. and {Venter}, C. and {Vialle}, J.~P. and {Viana}, A. and {Vincent}, P. and {Voelk}, H.~J. and {Volpe}, F. and {Vorobiov}, S. and {Vorster}, M. and {Wagner}, S.~J. and {Ward}, M. and {White}, R. and {Wierzcholska}, A. and {Zacharias}, M. and {Zajczyk}, A. and {Zdziarski}, A.~A. and {Zech}, A. and {Zechlin}, H. -S.~L.},
        title = "{Discovery of hard-spectrum {\ensuremath{\gamma}}-ray emission from the BL Lacertae object 1ES 0414+009}",
      journal = {\aap},
         year = 2012,
        month = feb,
       volume = {538},
          eid = {A103},
        pages = {A103},
          doi = {10.1051/0004-6361/201118406},
archivePrefix = {arXiv},
       eprint = {1201.2044},
 primaryClass = {astro-ph.HE},
       adsurl = {https://ui.adsabs.harvard.edu/abs/2012A&A...538A.103H}
}

@ARTICLE{2012ApJ...755..118A,
       author = {{Aliu}, E. and {Archambault}, S. and {Arlen}, T. and {Aune}, T. and {Beilicke}, M. and {Benbow}, W. and {B{\"o}ttcher}, M. and {Bouvier}, A. and {Bugaev}, V. and {Cannon}, A. and {Cesarini}, A. and {Ciupik}, L. and {Collins-Hughes}, E. and {Connolly}, M.~P. and {Cui}, W. and {Dickherber}, R. and {Dumm}, J. and {Errando}, M. and {Falcone}, A. and {Federici}, S. and {Feng}, Q. and {Finley}, J.~P. and {Finnegan}, G. and {Fortson}, L. and {Furniss}, A. and {Galante}, N. and {Gall}, D. and {Godambe}, S. and {Griffin}, S. and {Grube}, J. and {Gyuk}, G. and {Hanna}, D. and {Holder}, J. and {Huan}, H. and {Hughes}, G. and {Hui}, C.~M. and {Imran}, A. and {Jameil}, O. and {Kaaret}, P. and {Karlsson}, N. and {Kertzman}, M. and {Kerr}, J. and {Khassen}, Y. and {Kieda}, D. and {Krawczynski}, H. and {Krennrich}, F. and {Lang}, M.~J. and {Lee}, K. and {Madhavan}, A.~S. and {Majumdar}, P. and {McArthur}, S. and {McCann}, A. and {Moriarty}, P. and {Mukherjee}, R. and {Nelson}, T. and {O'Faol{\'a}in de Bhr{\'o}ithe}, A. and {Ong}, R.~A. and {Orr}, M. and {Otte}, A.~N. and {Park}, N. and {Perkins}, J.~S. and {Pichel}, A. and {Pohl}, M. and {Quinn}, J. and {Ragan}, K. and {Reynolds}, P.~T. and {Roache}, E. and {Ruppel}, J. and {Saxon}, D.~B. and {Schroedter}, M. and {Sembroski}, G.~H. and {{\c{S}}ent{\"u}rk}, G.~D. and {Smith}, A.~W. and {Staszak}, D. and {Stroh}, M. and {Telezhinsky}, I. and {Te{\v{s}}i{\'c}}, G. and {Theiling}, M. and {Thibadeau}, S. and {Tsurusaki}, K. and {Varlotta}, A. and {Vassiliev}, V.~V. and {Vivier}, M. and {Wakely}, S.~P. and {Ward}, J.~E. and {Weinstein}, A. and {Welsing}, R. and {Williams}, D.~A. and {Zitzer}, B.},
        title = "{Multiwavelength Observations of the AGN 1ES 0414+009 with VERITAS, Fermi-LAT, Swift-XRT, and MDM}",
      journal = {\apj},
         year = 2012,
        month = aug,
       volume = {755},
       number = {2},
          eid = {118},
        pages = {118},
          doi = {10.1088/0004-637X/755/2/118},
archivePrefix = {arXiv},
       eprint = {1206.4080},
 primaryClass = {astro-ph.HE},
       adsurl = {https://ui.adsabs.harvard.edu/abs/2012ApJ...755..118A}
}

@ARTICLE{1991AJ....101..818H,
       author = {{Halpern}, Jules P. and {Chen}, Vera S. and {Madejski}, Greg M. and {Chanan}, Gary A.},
        title = "{The Redshift of the X-ray Selected BL Lacertae Object H0414+009}",
      journal = {\aj},
         year = 1991,
        month = mar,
       volume = {101},
        pages = {818},
          doi = {10.1086/115725},
       adsurl = {https://ui.adsabs.harvard.edu/abs/1991AJ....101..818H}
}

@ARTICLE{2010A&A...516A..56H,
       author = {{H.~E.~S.~S. Collaboration} and {Abramowski}, A. and {Acero}, F. and {Aharonian}, F. and {Akhperjanian}, A.~G. and {Anton}, G. and {Barres de Almeida}, U. and {Bazer-Bachi}, A.~R. and {Becherini}, Y. and {Behera}, B. and {Benbow}, W. and {Bernl{\"o}hr}, K. and {Bochow}, A. and {Boisson}, C. and {Bolmont}, J. and {Borrel}, V. and {Brucker}, J. and {Brun}, F. and {Brun}, P. and {B{\"u}hler}, R. and {Bulik}, T. and {B{\"u}sching}, I. and {Boutelier}, T. and {Chadwick}, P.~M. and {Charbonnier}, A. and {Chaves}, R.~C.~G. and {Cheesebrough}, A. and {Conrad}, J. and {Chounet}, L. -M. and {Clapson}, A.~C. and {Coignet}, G. and {Costamante}, L. and {Dalton}, M. and {Daniel}, M.~K. and {Davids}, I.~D. and {Degrange}, B. and {Deil}, C. and {Dickinson}, H.~J. and {Djannati-Ata{\"\i}}, A. and {Domainko}, W. and {O'C. Drury}, L. and {Dubois}, F. and {Dubus}, G. and {Dyks}, J. and {Dyrda}, M. and {Egberts}, K. and {Eger}, P. and {Espigat}, P. and {Fallon}, L. and {Farnier}, C. and {Fegan}, S. and {Feinstein}, F. and {Fernandes}, M.~V. and {Fiasson}, A. and {F{\"o}rster}, A. and {Fontaine}, G. and {F{\"u}{\ss}ling}, M. and {Gabici}, S. and {Gallant}, Y.~A. and {G{\'e}rard}, L. and {Gerbig}, D. and {Giebels}, B. and {Glicenstein}, J.~F. and {Gl{\"u}ck}, B. and {Goret}, P. and {G{\"o}ring}, D. and {Hampf}, D. and {Hauser}, M. and {Heinz}, S. and {Heinzelmann}, G. and {Henri}, G. and {Hermann}, G. and {Hinton}, J.~A. and {Hoffmann}, A. and {Hofmann}, W. and {Hofverberg}, P. and {Holleran}, M. and {Hoppe}, S. and {Horns}, D. and {Jacholkowska}, A. and {de Jager}, O.~C. and {Jahn}, C. and {Jung}, I. and {Katarzy{\'n}ski}, K. and {Katz}, U. and {Kaufmann}, S. and {Kerschhaggl}, M. and {Khangulyan}, D. and {Kh{\'e}lifi}, B. and {Keogh}, D. and {Klochkov}, D. and {Klu{\v{z}}niak}, W. and {Kneiske}, T. and {Komin}, Nu. and {Kosack}, K. and {Kossakowski}, R. and {Lamanna}, G. and {Lenain}, J. -P. and {Lohse}, T. and {Lu}, C. -C. and {Marandon}, V. and {Marcowith}, A. and {Masbou}, J. and {Maurin}, D. and {McComb}, T.~J.~L. and {Medina}, M.~C. and {M{\'e}hault}, J. and {Moderski}, R. and {Moulin}, E. and {Naumann-Godo}, M. and {de Naurois}, M. and {Nedbal}, D. and {Nekrassov}, D. and {Nguyen}, N. and {Nicholas}, B. and {Niemiec}, J. and {Nolan}, S.~J. and {Ohm}, S. and {Olive}, J. -F. and {de O{\~n}a Wilhelmi}, E. and {Opitz}, B. and {Orford}, K.~J. and {Ostrowski}, M. and {Panter}, M. and {Paz Arribas}, M. and {Pedaletti}, G. and {Pelletier}, G. and {Petrucci}, P. -O. and {Pita}, S. and {P{\"u}hlhofer}, G. and {Punch}, M. and {Quirrenbach}, A. and {Raubenheimer}, B.~C. and {Raue}, M. and {Rayner}, S.~M. and {Reimer}, O. and {Renaud}, M. and {de Los Reyes}, R. and {Rieger}, F. and {Ripken}, J. and {Rob}, L. and {Rosier-Lees}, S. and {Rowell}, G. and {Rudak}, B. and {Rulten}, C.~B. and {Ruppel}, J. and {Ryde}, F. and {Sahakian}, V. and {Santangelo}, A. and {Schlickeiser}, R. and {Sch{\"o}ck}, F.~M. and {Sch{\"o}nwald}, A. and {Schwanke}, U. and {Schwarzburg}, S. and {Schwemmer}, S. and {Shalchi}, A. and {Sushch}, I. and {Sikora}, M. and {Skilton}, J.~L. and {Sol}, H. and {Stawarz}, {\L}. and {Steenkamp}, R. and {Stegmann}, C. and {Stinzing}, F. and {Szostek}, A. and {Tam}, P.~H. and {Tavernet}, J. -P. and {Terrier}, R. and {Tibolla}, O. and {Tluczykont}, M. and {Valerius}, K. and {van Eldik}, C. and {Vasileiadis}, G. and {Venter}, C. and {Venter}, L. and {Vialle}, J.~P. and {Viana}, A. and {Vincent}, P. and {Vivier}, M. and {V{\"o}lk}, H.~J. and {Volpe}, F. and {Vorobiov}, S. and {Wagner}, S.~J. and {Ward}, M. and {Zdziarski}, A.~A. and {Zech}, A. and {Zechlin}, H. -S.},
        title = "{Multi-wavelength observations of H 2356-309}",
      journal = {\aap},
         year = 2010,
        month = jun,
       volume = {516},
          eid = {A56},
        pages = {A56},
          doi = {10.1051/0004-6361/201014321},
archivePrefix = {arXiv},
       eprint = {1004.2089},
 primaryClass = {astro-ph.HE},
       adsurl = {https://ui.adsabs.harvard.edu/abs/2010A&A...516A..56H}
}

@ARTICLE{2015EPJC...75..273S,
       author = {{Sahu}, Sarira and {Miranda}, Luis Salvador},
        title = "{Some possible sources of IceCube TeV{\textendash}PeV neutrino events}",
      journal = {European Physical Journal C},
         year = 2015,
        month = jun,
       volume = {75},
          eid = {273},
        pages = {273},
          doi = {10.1140/epjc/s10052-015-3519-1},
archivePrefix = {arXiv},
       eprint = {1408.3664},
 primaryClass = {astro-ph.HE},
       adsurl = {https://ui.adsabs.harvard.edu/abs/2015EPJC...75..273S}
}

@ARTICLE{2007A&A...470..475A,
       author = {{Aharonian}, F. and {Akhperjanian}, A.~G. and {Bazer-Bachi}, A.~R. and {Beilicke}, M. and {Benbow}, W. and {Berge}, D. and {Bernl{\"o}hr}, K. and {Boisson}, C. and {Bolz}, O. and {Borrel}, V. and {Braun}, I. and {Brion}, E. and {Brown}, A.~M. and {B{\"u}hler}, R. and {B{\"u}sching}, I. and {Boutelier}, T. and {Carrigan}, S. and {Chadwick}, P.~M. and {Chounet}, L. -M. and {Coignet}, G. and {Cornils}, R. and {Costamante}, L. and {Degrange}, B. and {Dickinson}, H.~J. and {Djannati-Ata{\"\i}}, A. and {O'C. Drury}, L. and {Dubus}, G. and {Egberts}, K. and {Emmanoulopoulos}, D. and {Espigat}, P. and {Farnier}, C. and {Feinstein}, F. and {Ferrero}, E. and {Fiasson}, A. and {Fontaine}, G. and {Funk}, Seb. and {Funk}, S. and {F{\"u}{\ss}ling}, M. and {Gallant}, Y.~A. and {Giebels}, B. and {Glicenstein}, J.~F. and {Gl{\"u}ck}, B. and {Goret}, P. and {Hadjichristidis}, C. and {Hauser}, D. and {Hauser}, M. and {Heinzelmann}, G. and {Henri}, G. and {Hermann}, G. and {Hinton}, J.~A. and {Hoffmann}, A. and {Hofmann}, W. and {Holleran}, M. and {Hoppe}, S. and {Horns}, D. and {Jacholkowska}, A. and {de Jager}, O.~C. and {Kendziorra}, E. and {Kerschhaggl}, M. and {Kh{\'e}lifi}, B. and {Komin}, Nu. and {Kosack}, K. and {Lamanna}, G. and {Latham}, I.~J. and {Le Gallou}, R. and {Lemi{\`e}re}, A. and {Lemoine-Goumard}, M. and {Lohse}, T. and {Martin}, J.~M. and {Martineau-Huynh}, O. and {Marcowith}, A. and {Masterson}, C. and {Maurin}, G. and {McComb}, T.~J.~L. and {Moulin}, E. and {de Naurois}, M. and {Nedbal}, D. and {Nolan}, S.~J. and {Noutsos}, A. and {Olive}, J. -P. and {Orford}, K.~J. and {Osborne}, J.~L. and {Panter}, M. and {Pelletier}, G. and {Petrucci}, P. -O. and {Pita}, S. and {P{\"u}hlhofer}, G. and {Punch}, M. and {Ranchon}, S. and {Raubenheimer}, B.~C. and {Raue}, M. and {Rayner}, S.~M. and {Ripken}, J. and {Rob}, L. and {Rolland}, L. and {Rosier-Lees}, S. and {Rowell}, G. and {Sahakian}, V. and {Santangelo}, A. and {Saug{\'e}}, L. and {Schlenker}, S. and {Schlickeiser}, R. and {Schr{\"o}der}, R. and {Schwanke}, U. and {Schwarzburg}, S. and {Schwemmer}, S. and {Shalchi}, A. and {Sol}, H. and {Spangler}, D. and {Spanier}, F. and {Steenkamp}, R. and {Stegmann}, C. and {Superina}, G. and {Tam}, P.~H. and {Tavernet}, J. -P. and {Terrier}, R. and {Tluczykont}, M. and {van Eldik}, C. and {Vasileiadis}, G. and {Venter}, C. and {Vialle}, J.~P. and {Vincent}, P. and {V{\"o}lk}, H.~J. and {Wagner}, S.~J. and {Ward}, M.},
        title = "{Detection of VHE gamma-ray emission from the distant blazar 1ES{\,}1101-232 with HESS and broadband characterisation}",
      journal = {\aap},
         year = 2007,
        month = aug,
       volume = {470},
       number = {2},
        pages = {475-489},
          doi = {10.1051/0004-6361:20077057},
archivePrefix = {arXiv},
       eprint = {0705.2946},
 primaryClass = {astro-ph},
       adsurl = {https://ui.adsabs.harvard.edu/abs/2007A&A...470..475A}
}

@ARTICLE{1994ApJS...93..125F,
       author = {{Falomo}, R. and {Scarpa}, R. and {Bersanelli}, M.},
        title = "{Optical Spectrophotometry of Blazars}",
      journal = {\apjs},
         year = 1994,
        month = jul,
       volume = {93},
        pages = {125},
          doi = {10.1086/192048},
       adsurl = {https://ui.adsabs.harvard.edu/abs/1994ApJS...93..125F}
}

@ARTICLE{2007Ap&SS.309..487C,
       author = {{Costamante}, Luigi},
        title = "{A low density of the extragalactic background light revealed by the H.E.S.S. spectra of the BL Lac objects 1ES 1101-232 and H 2356-309}",
      journal = {\apss},
         year = 2007,
        month = jun,
       volume = {309},
       number = {1-4},
        pages = {487-495},
          doi = {10.1007/s10509-007-9418-7},
archivePrefix = {arXiv},
       eprint = {astro-ph/0612709},
 primaryClass = {astro-ph},
       adsurl = {https://ui.adsabs.harvard.edu/abs/2007Ap&SS.309..487C}
}

@ARTICLE{2007A&A...473L..25A,
       author = {{Aharonian}, F. and {Akhperjanian}, A.~G. and {Barres de Almeida}, U. and {Bazer-Bachi}, A.~R. and {Behera}, B. and {Beilicke}, M. and {Benbow}, W. and {Bernl{\"o}hr}, K. and {Boisson}, C. and {Bolz}, O. and {Borrel}, V. and {Braun}, I. and {Brion}, E. and {Brown}, A.~M. and {B{\"u}hler}, R. and {Bulik}, T. and {B{\"u}sching}, I. and {Boutelier}, T. and {Carrigan}, S. and {Chadwick}, P.~M. and {Chounet}, L. -M. and {Clapson}, A.~C. and {Coignet}, G. and {Cornils}, R. and {Costamante}, L. and {Dalton}, M. and {Degrange}, B. and {Dickinson}, H.~J. and {Djannati-Ata{\"\i}}, A. and {Domainko}, W. and {O'C. Drury}, L. and {Dubois}, F. and {Dubus}, G. and {Dyks}, J. and {Egberts}, K. and {Emmanoulopoulos}, D. and {Espigat}, P. and {Farnier}, C. and {Feinstein}, F. and {Fiasson}, A. and {F{\"o}rster}, A. and {Fontaine}, G. and {Funk}, Seb. and {F{\"u}{\ss}ling}, M. and {Gallant}, Y.~A. and {Giebels}, B. and {Glicenstein}, J.~F. and {Gl{\"u}ck}, B. and {Goret}, P. and {Hadjichristidis}, C. and {Hauser}, D. and {Hauser}, M. and {Heinzelmann}, G. and {Henri}, G. and {Hermann}, G. and {Hinton}, J.~A. and {Hoffmann}, A. and {Hofmann}, W. and {Holleran}, M. and {Hoppe}, S. and {Horns}, D. and {Jacholkowska}, A. and {de Jager}, O.~C. and {Jung}, I. and {Katarzy{\'n}ski}, K. and {Kendziorra}, E. and {Kerschhaggl}, M. and {Kh{\'e}lifi}, B. and {Keogh}, D. and {Komin}, Nu. and {Kosack}, K. and {Lamanna}, G. and {Latham}, I.~J. and {Lemi{\`e}re}, A. and {Lemoine-Goumard}, M. and {Lenain}, J. -P. and {Lohse}, T. and {Martin}, J.~M. and {Martineau-Huynh}, O. and {Marcowith}, A. and {Masterson}, C. and {Maurin}, D. and {Maurin}, G. and {McComb}, T.~J.~L. and {Moderski}, R. and {Moulin}, E. and {de Naurois}, M. and {Nedbal}, D. and {Nolan}, S.~J. and {Ohm}, S. and {Olive}, J. -P. and {de O{\~n}a Wilhelmi}, E. and {Orford}, K.~J. and {Osborne}, J.~L. and {Ostrowski}, M. and {Panter}, M. and {Pedaletti}, G. and {Pelletier}, G. and {Petrucci}, P. -O. and {Pita}, S. and {P{\"u}hlhofer}, G. and {Punch}, M. and {Ranchon}, S. and {Raubenheimer}, B.~C. and {Raue}, M. and {Rayner}, S.~M. and {Renaud}, M. and {Ripken}, J. and {Rob}, L. and {Rolland}, L. and {Rosier-Lees}, S. and {Rowell}, G. and {Rudak}, B. and {Ruppel}, J. and {Sahakian}, V. and {Santangelo}, A. and {Schlickeiser}, R. and {Sch{\"o}ck}, F. and {Schr{\"o}der}, R. and {Schwanke}, U. and {Schwarzburg}, S. and {Schwemmer}, S. and {Shalchi}, A. and {Sol}, H. and {Spangler}, D. and {Stawarz}, {\L}. and {Steenkamp}, R. and {Stegmann}, C. and {Superina}, G. and {Tam}, P.~H. and {Tavernet}, J. -P. and {Terrier}, R. and {van Eldik}, C. and {Vasileiadis}, G. and {Venter}, C. and {Vialle}, J.~P. and {Vincent}, P. and {Vivier}, M. and {V{\"o}lk}, H.~J. and {Volpe}, F. and {Wagner}, S.~J. and {Ward}, M. and {Zdziarski}, A.~A. and {Zech}, A.},
        title = "{Discovery of VHE {\ensuremath{\gamma}}-rays from the distant BL Lacertae 1ES 0347-121}",
      journal = {\aap},
         year = 2007,
        month = oct,
       volume = {473},
       number = {3},
        pages = {L25-L28},
          doi = {10.1051/0004-6361:20078412},
archivePrefix = {arXiv},
       eprint = {0708.3021},
 primaryClass = {astro-ph},
       adsurl = {https://ui.adsabs.harvard.edu/abs/2007A&A...473L..25A}
}

@ARTICLE{2009MNRAS.396L.105G,
       author = {{Ghisellini}, G. and {Maraschi}, L. and {Tavecchio}, F.},
        title = "{The Fermi blazars' divide}",
      journal = {\mnras},
         year = 2009,
        month = jun,
       volume = {396},
       number = {1},
        pages = {L105-L109},
          doi = {10.1111/j.1745-3933.2009.00673.x},
archivePrefix = {arXiv},
       eprint = {0903.2043},
 primaryClass = {astro-ph.CO},
       adsurl = {https://ui.adsabs.harvard.edu/abs/2009MNRAS.396L.105G}
}

@BOOK{1992apa..book.....F,
       author = {{Frank}, J. and {King}, A. and {Raine}, D.},
        title = "{Accretion power in astrophysics.}",
         year = 1992,
       volume = {21},
       adsurl = {https://ui.adsabs.harvard.edu/abs/1992apa..book.....F}
}

@ARTICLE{2000ApJ...537..236Y,
       author = {{Yuan}, Feng and {Peng}, Qiuhe and {Lu}, Ju-fu and {Wang}, Jianmin},
        title = "{The Role of the Outer Boundary Condition in Accretion Disk Models: Theory and Application}",
      journal = {\apj},
         year = 2000,
        month = jul,
       volume = {537},
       number = {1},
        pages = {236-244},
          doi = {10.1086/309020},
archivePrefix = {arXiv},
       eprint = {astro-ph/0002068},
 primaryClass = {astro-ph},
       adsurl = {https://ui.adsabs.harvard.edu/abs/2000ApJ...537..236Y}
}
\bibliographystyle{aasjournal}
\clearpage

\begin{table*}
\setlength{\tabcolsep}{5.5pt}  % 默认是 6pt
\renewcommand{\arraystretch}{1.1} % 设置行间距
\begin{center}
\caption{SED Fitting Parameters}
\label{table_SED}
{\footnotesize %\scriptsize 
\begin{tabular}{lcccc cr} 
\toprule
\toprule
\multirow{2}{*}{Parameter} & \multirow{2}{*}{Symbol}   &\multirow{2}{*}{1ES 0229+200}&\multirow{2}{*}{H2356--309}&\multirow{2}{*}{1ES 1101--232}&\multirow{2}{*}{1ES 0347--121} &\multirow{2}{*}{1ES 0414+009} \\
&	   	&   \\
\hline
Redshift                     & $z$                                & $0.139$         &$0.165$            &$0.186$           &$0.188$        &$0.287$\\
\hline
\rowcolor{gray!25} \multicolumn{7}{c}{Leptonic Jet Model} \\  
\hline
Electron Density Parameter          & $N_{\rm e,0}$ [cm$^{-3}$]  & $7\times10^{1}$ &$6\times10^{2}$ &$2.5\times10^{1}$&$9\times10^{3}$ &$90$\\
Spectral Index below Break           & $p_1$                           & $1.85$          &$2.15$             &$1.8$             &$2.4$             &$2.15$    \\
Spectral Index above Break           & $p_2$                            & $3.16$          & $3.3$             &$3.4$             &$3$             & $3.55$    \\
Minimum Lorentz Factor for Electrons &$\gamma_{\rm e,min}$               & $1$             &$2$                &$5$               &$1\times10^{2}$             &$90$        \\
Break Lorentz Factor for Electrons   &$\gamma_{\rm e,b}$    & $7\times10^{5}$ &$1.4\times10^{5}$  &$7.3\times10^{4}$   &$2\times10^{5}$ &$7.6\times10^{4}$\\
Maximum Lorentz Factor for Electrons &$\gamma_{\rm e,max}$  & $ 7\times10^{6}$&$1.4\times10^{6}$  &$7.3\times10^{5}$   &$2\times10^{6}$ & $1.5\times10^{5}$\\
$^{\star}$Variability Timescale      & $\Delta t$ [hr]                    & $24$            & $24$              & $12$             & $12 $            &$62.5$\\		
Doppler Factor                       & $\delta_{\rm D}$                 & $8$             &$20$               &$20$              &$20$              &$30$\\
Magnetic Field Strength              & $B_{\rm jet}$ [G]               & $0.55$          &$0.04$            &$0.25$             &$0.2$            &$0.025$      \\
\hline
\rowcolor{gray!25} \multicolumn{7}{c}{Lepto-hadronic ADAF Model} \\ 
\hline 
$^{\star}$ Mass of SMBH              & $M_{\rm BH}$ [$M_{\odot}$]& $1.75\times10^{9}$  &$3.98\times10^{8}$&$1.02\times10^{9}$&$4.47\times10^{8}$&$2\times10^{9}$\\
$^{\star}$ Accretion Rate            &$\dot{m}$   &$4.25\times10^{-4}$&$3\times10^{-3}$&$2.7\times10^{-3}$&$6.42\times10^{-3}$& $7.3\times10^{-3}$\\
$^{\star}$ Power-law Index for Radial Variation  &$s$      & 0.3  &0.3&0.3&0.3&0.3\\
$^{\star}$ Shakura-Sunyaev Viscosity Parameter   &$\alpha$     & 0.3  &0.3&0.3&0.3&0.3\\
$^{\star}$ Ratio between the Gas and the Total Pressure  &$\beta$     & 0.9  &0.9&0.9&0.9&0.9\\
$^{\star}$ Adiabatic Index                    &$\gamma_{\rm adi}$     & 1.5  &1.5&1.5&1.5&1.5\\
$^{\star}$ Outer Radius of the ADAF           & $R_{\rm o}$ [$R_{\rm S}$] & $10^{4}$  &$10^{4}$&$10^{4}$&$10^{4}$&$10^{4}$\\	
Fraction of Energy to Electrons     &$\delta_{\rm ADAF}$     & 0.1  &0.5&0.4&0.5&0.5\\
Differential Proton Density at 1 eV  & $N_{\rm p,0}\ $[$\rm cm^{-3}\ eV^{-1}$] & $2.5\times10^{13}$ &$2.5\times10^{11}$&$1.6\times10^{12}$ &$1.5\times10^{11}$&$3\times10^{11}$\\
Integrated Proton Density            & $N_{\rm p}\ $[$\rm cm^{-3}$]     & $1.1\times10^{7}$ &$1.7\times10^{6}$&$6.4\times10^{4}$ &$1.1\times10^{5}$&$1.9\times10^{4}$\\
Proton Spectral Index             & $p$                & $1.6$    &$1.6$ &$1.7$ &$1.6$ &$1.7$  \\
Proton Cutoff Energy              & $\varepsilon_{\rm p, cut}$ [eV]   & $9\times10^{13}$  &$3\times10^{13}$&$6\times10^{13}$ &$4\times10^{13}$&$3\times10^{13}$\\
$^{\star}$Average Magnetic Field of ADAF    & $B_{\rm ADAF}$ [G]         & $1 $   &$1 $&$1 $&$1 $&$1 $\\
\hline          
\rowcolor{gray!25} \multicolumn{7}{c}{the $pp$ Process} \\  
\hline
Cold Proton Number Density & $n_{\rm p}$ [cm$^{-3}$]  & $1\times10^{-4} $   &$1.5$&$0.4 $&$3 $&$0.5 $\\
\bottomrule
\end{tabular}
}   
\end{center}
\textbf{Notes.} 
The values of $M_{\rm BH}$ for 1ES 0229+200, H2356--309, and 1ES 0347--121 are taken from \citet{2002ApJ...579..530W}, while those for 1ES 0414+009 and 1ES 1101--232 are adopted from \citet{2003MNRAS.343..505F} and \citet{2021ApJS..253...46P}, respectively.\\
$^\star$ It indicates that the parameter values are fixed.
\end{table*}

% \footnote{\url{http://www.iasf-milano.inaf.it/~polletta/templates/swire_templates.html}}

% \protect

\end{document}